\documentclass[journal]{IEEEtran}
\ifCLASSINFOpdf
\else
\fi

\usepackage{ifpdf}
\usepackage[pdftex]{graphicx}
\usepackage[cmex10]{amsmath}
\usepackage{subfig}
\usepackage{amssymb}
\usepackage{mathrsfs}
\usepackage{amsmath}
\usepackage{indentfirst}
\usepackage{booktabs}
\usepackage{multirow}
\usepackage{graphicx}
\usepackage{epstopdf}
\usepackage{fixltx2e}
\usepackage{soul}
\usepackage{units}
\usepackage{mathtools}
\usepackage{framed}

\usepackage[utf8]{inputenc}
\usepackage{url} 
\usepackage{cite} 
\usepackage[numbers]{natbib}

\usepackage{caption}
\usepackage{epsfig,graphicx,amssymb,amsmath}
\usepackage{subfig}
\usepackage{diagbox}
\usepackage{color}
\usepackage{setspace}

\usepackage{amsthm}
\usepackage{algorithm}
\usepackage[noend]{algpseudocode}
\algrenewcommand\algorithmicend{\textbf{end}}
\algrenewcommand\algorithmicif{\textbf{if}}
\algrenewcommand\algorithmicfor{\textbf{for}}
\usepackage{booktabs}
\usepackage{diagbox}
\usepackage{float}
\usepackage{epstopdf}
\usepackage{ragged2e}
\renewcommand{\raggedright}{\leftskip=0pt \rightskip=0pt plus 0cm}

\ifCLASSINFOpdf
\else

\fi

\usepackage{xcolor}

\usepackage{xpatch}

\begin{document}
%
\title{Multi-AUV Cooperative Target Tracking Based on Supervised Diffusion-Aided Multi-Agent Reinforcement Learning}

\author{Jiaao Ma,
Chuan Lin,~\IEEEmembership{Member,~IEEE},
Guangjie Han,~\IEEEmembership{Fellow,~IEEE},
Shengchao~Zhu,~\IEEEmembership{Student Member,~IEEE,}
Zhenyu Wang,
Chen An

\thanks{\emph{Corresponding author: Guangjie Han}} 
\thanks{Jiaao Ma, Chuan Lin, and Zhenyu Wang are with Software College, Northeastern University, Shenyang, China. (e-mails: 2727746375@qq.com; chuanlin1988@gmail.com; larrywang1019@outlook.com;).}
\thanks{Guangjie Han is with Key Laboratory of Maritime Intelligent Network Information Technology, Ministry of Education, Hohai University. (e-mails: hanguangjie@gmail.com).}
\thanks{Shengchao Zhu is with the College of Computer Science and Software Engineering, Hohai
University, Nanjing, 210013, China (e-mail: zhushengchao77@gmail.com).}
\thanks{Chen An is with the College of Computer Science and Engineering, Northeastern University, Shenyang, China (e-mail:  20235854@stu.neu.edu.cn).}

 	
}

\markboth{Journal of \LaTeX\ Class Files,~Vol.~XX, No.~X, XXX~XXXX}%
{Shell \MakeLowercase{\textit{et al.}}: Bare Demo of IEEEtran.cls for IEEE Journals}
%

\IEEEtitleabstractindextext{%
	\begin{abstract}
	
In recent years, advances in underwater networking and multi-agent reinforcement learning (MARL) have significantly expanded multi-autonomous underwater vehicle (AUV) applications in marine exploration and target tracking. 
However, current MARL-driven cooperative tracking faces three critical challenges: 1) non-stationarity in decentralized coordination, where local policy updates destabilize teammates' observation spaces, preventing convergence; 2) sparse-reward exploration inefficiency from limited underwater visibility and constrained sensor ranges, causing high-variance learning; and 3) water disturbance fragility combined with handcrafted reward dependency that degrades real-world robustness under unmodeled hydrodynamic conditions. 
To address these challenges, this paper proposes a hierarchical MARL architecture comprising four layers: global training scheduling, multi-agent coordination, local decision-making, and real-time execution. This architecture optimizes task allocation and inter-AUV coordination through hierarchical decomposition. 
Building on this foundation, we propose the Supervised Diffusion-Aided MARL (SDA-MARL) algorithm featuring three innovations: 1) a dual-decision architecture with segregated experience pools mitigating non-stationarity through structured experience replay; 2) a supervised learning mechanism guiding the diffusion model's reverse denoising process to generate high-fidelity training samples that accelerate convergence; and 3) disturbance-robust policy learning incorporating behavioral cloning loss to guide the Deep Deterministic Policy Gradient (DDPG) network update using high-quality replay actions, eliminating handcrafted reward dependency.
The tracking algorithm based on SDA-MARL proposed in this paper achieves superior precision, robustness to dynamic ocean conditions, and computational efficiency compared to state-of-the-art methods in comprehensive underwater simulations.

		
\end{abstract}
	
	}

\maketitle
\IEEEdisplaynontitleabstractindextext
\IEEEpeerreviewmaketitle

\IEEEpeerreviewmaketitle

\section{Introduction}

\label{sec:introduction}

\IEEEPARstart{A}{}s the largest connected ecosystem on Earth, the ocean serves as a critical arena for resource exploitation, environmental monitoring, and maritime security \cite{Hoegh-Guldberg2023}. 
To meet these demands, multi-AUV systems have been extensively developed and widely applied to seabed resource exploration, ecological monitoring, scientific surveys, and especially cooperative tracking tasks, where multiple AUVs must jointly pursue dynamic targets and maintain coordinated following of underwater objects \cite{Wang2023AnAC,9356608}. 
However, the complexity of the marine environment and the dynamic nature of underwater target motion pose substantial challenges to reliable and efficient cooperative tracking \cite{4089061}.

Multi-agent reinforcement learning (MARL) offers a distributed decision-making framework to coordinate multiple AUVs, enabling agents to learn cooperative strategies through environmental interaction while balancing global objectives with individual goals \cite{10675327}. 
Compared to centralized control, MARL offers key advantages, including decentralized perception, asynchronous decision-making, and strong adaptability to environmental uncertainties, making it particularly suitable for multi-AUV systems where each AUV acts as an autonomous agent that collaborates through local observations \cite{9685323}. 
Recent studies have reported notable progress in the application of MARL to multi-AUV cooperation, particularly in tracking dynamic underwater targets under the centralized training paradigm with decentralized execution (CTDE) \cite{10.5555/3237383.3238080}.

Despite these advances, existing learning-based methods for underwater cooperative tracking remain limited. 
Conventional single-agent RL or centralized RL frameworks suffer from poor scalability and heavy communication burdens when extended to multi-AUV scenarios. 
Moreover, many methods rely on accurate environmental modeling and handcrafted reward functions, rendering them fragile to unmodeled hydrodynamic disturbances and domain shifts in real-world ocean environments \cite{Paulius-RSS-20}. 
Even within MARL frameworks, the combination of partial observability, delayed sparse rewards, and strong environmental disturbances frequently leads to unstable training and policies that are difficult to deploy safely.

However, current MARL-driven cooperative tracking in multi-AUV systems faces three fundamental challenges:
1) non-stationarity in decentralized coordination: Local policy updates in decentralized settings dynamically alter teammates' observation spaces, inducing non-stationary environments that prevent stable policy convergence \cite{pmlr-v232-nekoei23a,10.1007/978-3-031-20309-1_35};
2) sparse-reward exploration inefficiency: Limited underwater visibility and constrained sensor ranges yield sparse delayed rewards during target tracking, causing high-variance exploration and slow learning \cite{pmlr-v80-rashid18a,jmse10101406};
3) unmodeled hydrodynamic disturbances and reward dependency: Unmodeled hydrodynamic disturbances combined with handcrafted reward functions critically degrade policy robustness and real-world deployment reliability \cite{Paulius-RSS-20,machines13060503}.

To address the aforementioned issues, this paper introduces the concept of efficient sample generation and proposes a Supervised Diffusion-Aided MARL (SDA-MARL) framework for multi-AUV cooperative tracking. The framework integrates three core innovations:
1) a dual-decision architecture that mitigates non-stationarity through segregated experience pools;
2) a supervised learning mechanism that leverages high-quality replay actions to accelerate convergence and reduce exploration variance;
3) disturbance-robust policy learning via a diffusion model trained on high-quality replay actions to generate robust actions, eliminating dependency on handcrafted rewards.
In total, the main contributions of this work are:

\begin{enumerate}[]
    \item Based on diffusion models, we propose SDA-MARL, an algorithm that schedules multi-AUV systems for target tracking in sparse-reward environments. Within SDA-MARL, we design a dual-decision network architecture integrated with a structured experience replay mechanism to mitigate non-stationarity during MARL training.
    
    \item Furthermore, in SDA-MARL, we introduce a supervised learning component to guide the update of the diffusion model. By training the diffusion model to match high-quality actions from the replay buffer, this supervised signal reduces exploratory variance during early training stages and accelerates convergence, while also maintaining stability through direct imitation of demonstrated behaviors.

    \item Meanwhile, in SDA-MARL, we introduce a behavioral cloning loss to guide the training of the Deep Deterministic Policy Gradient (DDPG) Actor network. This loss trains the Actor to imitate high-quality actions synthesized by the diffusion model based on the replay buffer, which accelerates policy convergence and enhances robustness to dynamic underwater disturbances, thereby reducing the dependence on explicit reward engineering.
\end{enumerate}

The remainder of this paper is organized as follows. Section \ref{Section:2} reviews the related work. Section \ref{Section:3} displays the formulation of the problem and the system preliminaries. Section \ref{Section:4} proposes the SDA-MARL framework. Section \ref{Section:5} details the SDA-MARL-based cooperative tracking algorithm for multi-AUV systems. Section \ref{Section:6} presents experimental evaluations and results. Finally, Section \ref{Section:7} concludes the paper and outlines future research directions.

\section{Related Works}\label{Section:2}
In this section, we mainly study the latest research products related to the subject, specifically divided into the following two areas: 1) AUV target tracking algorithms based on RL/MARL; 2) reinforcement learning methods based on diffusion models.

\subsection{AUV Target Tracking Algorithms Based on RL/MARL}\label{Section:2-1}
Existing research demonstrates that single-AUV target tracking systems provide integrated perception, planning, and control capabilities \cite{ELHAKI2018239}, delivering reliable performance in well-defined small-scale scenarios through their simple architecture. 
However, these systems exhibit inherent limitations in complex, large-scale, dynamic multi-target environments: constrained perception range, insufficient parallel tasking capability, poor disturbance resistance, and robustness overly dependent on a single platform \cite{11125482}.

To overcome these bottlenecks, the research focus has shifted to multi-AUV cooperative tracking systems. 
Utilizing swarm control and distributed coordination, such systems enable real-time information sharing and joint decision-making among platforms. 
This significantly extends the perceptual field, enhances parallel multi-target tracking capability, and improves overall system resilience and mission efficiency \cite{9938376}.

In \cite{HADI2022103326}, the authors propose a TD3 deep reinforcement learning-based end-to-end motion planning and control strategy for AUVs. Using a 6-DOF dynamic model with direct rudder angle control, the method achieves robust path planning and precise target arrival in unknown environments with ocean currents through a composite reward function.

In \cite{WU2025122855}, the authors present a few-shot underwater visual object tracking framework integrating a Kalman filter with reinforcement learning-based twin network optimization. 
Its hybrid architecture coordinates motion prior and dynamic hyperparameter optimization to achieve robust tracking under data scarcity and complex underwater disturbances, validated on experimental platforms.

In \cite{10650551}, the authors develop a multi-AUV cooperative perception and decision-making framework for dynamic target tracking. 
The hybrid architecture combines CNN-based target localization using axis-frequency electric field signals with MARL, modeling cooperative tracking as a partially observable Markov game to enable autonomous decision-making under constraints in unknown environments.

While RL/MARL has been widely used in multi-AUV tracking tasks, it still struggles with environmental adaptability, training stability, and multi-target robustness. Diffusion models, with their strong policy generation capabilities, now offer a promising way to build more scalable and robust collaborative decision-making frameworks for multi-AUV systems.

\subsection{Reinforcement Learning Methods Based on Diffusion Models}\label{Section:2-2}
Recently, diffusion model-driven reinforcement learning has emerged as a promising interdisciplinary frontier, bridging the gap between generative AI and decision-making intelligence \cite{NEURIPS2023_ccda3c63}. 
Leveraging probabilistic foundations and strong generative capabilities, the diffusion model effectively supports core tasks including policy optimization, environment modeling, and multi-agent collaboration \cite {Xian2023ChainedDiffuserUT}. 
This has spurred diverse studies that underscore its potential to boost policy sampling efficiency, enhance representational diversity and robustness, and facilitate multi-agent coordination—offering a promising pathway toward high-performance decision systems in complex dynamic environments \cite{10.5555/3737916.3738052}.

In \cite{wang2023diffusion}, the authors propose Diffusion-QL, an offline reinforcement learning algorithm that employs a state-conditioned diffusion model to represent multi-modal policies. 
The method injects Q-learning gradients directly into the diffusion reverse process, guiding action generation toward high-value regions while respecting data constraints, and achieves state-of-the-art performance on D4RL benchmarks.

In \cite{zheng2024safe}, the authors propose FISOR, a framework that integrates reachability analysis with diffusion models to enforce hard safety constraints in offline RL. It identifies the maximum feasible region via inverse expectile regression, training a feasibility-aware policy through weighted diffusion to maximize rewards in feasible states and minimize violations elsewhere. On DSRL benchmarks, FISOR uniquely satisfies all safety requirements across tasks while achieving state-of-the-art returns.

In \cite{ding2024diffusionbased}, the authors propose QVPO, a method that integrates diffusion models into online reinforcement learning. The approach utilizes a Q-weighted variational loss for policy optimization, uniform-sample entropy regularization for enhanced exploration, and a behavior policy that selects high-value actions from multiple diffusion samples. This method achieves state-of-the-art performance on MuJoCo benchmarks, outperforming baseline algorithms including SAC, TD3, DIPO, and QSM.

In summary, despite progress in multi-AUV cooperative tracking and diffusion-based RL, critical challenges persist in dynamic environment adaptation, swarm training stability, and robust multi-target tracking. To bridge these gaps, this paper proposes a hierarchical generative MARL framework guided by a supervised diffusion technique, enhancing scalability and robustness for complex underwater operations.

\section{Preliminary Materials}\label{Section:3}
In this section, to accurately simulate the process of avoiding obstacles and tracking targets underwater while adapting to real dynamic effects, this paper incorporates the influence of ocean currents into the modeling. 
Furthermore, we transform the tracking problem in multi-AUV systems into a Markov decision process (MDP).

\subsection{Modeling of Underwater Environment Detection}\label{Section:3-1}

In complex underwater environments, accurately determining the relative positions between AUVs and their targets is critical. 
In this work, the sonar technology is employed for AUV positioning. 
A sonar-equipped AUV emits acoustic waves to scan its surroundings. A fan-shaped array then captures echo signals from various directions, enabling target localization based on echo strength. 
The entire detection process is mathematically modeled and optimized using the classic active sonar equation presented as Eq. \eqref{eq1}:

\begin{equation}
\label{eq1}
\mathrm{EM} = \mathrm{SL} - 2\mathrm{TL} + \mathrm{TS} - (\mathrm{NL} - \mathrm{DI}) - \mathrm{DT},
\end{equation}
where \(\mathrm{EM}\) is excess margin, \(\mathrm{SL}\) is source level, \(\mathrm{TL}\) is transmission loss, \(\mathrm{TS}\) is target strength, \(\mathrm{NL}\) is ambient noise level, \(\mathrm{DI}\) is the directional index, \(\mathrm{DT}\) is the detection threshold (all in dB).

\subsection{Ocean Circulation Modeling}\label{Section:3-2}

The complexity of the marine environment, particularly the dynamics of ocean currents, presents significant challenges for AUV-based tracking. 
To quantify and simulate the fluid dynamic forces central to these environmental interactions, the fundamental Navier-Stokes equations are employed, as given in Eq. \eqref{eq2}:

\begin{equation}
\label{eq2}
\rho \left( \frac{\partial \mathrm{u}}{\partial t} + \mathrm{u} \cdot \nabla \mathrm{u} \right) = -\nabla p + \mu \nabla^2 \mathrm{u} + \mathrm{F},
\end{equation}
where \(\rho\) is fluid density, \(\mathrm{u}\) is velocity field, \(\frac{\partial \mathrm{u}}{\partial t}\) is temporal derivative, \(\mathrm{u} \cdot \nabla \mathrm{u}\) is convective term, \(\nabla p\) is pressure gradient, \(\mu\) is dynamic viscosity, \(\nabla^2 \mathrm{u}\) is diffusive term, and \(\mathrm{F}\) is external force.

Meanwhile, our work models as well as the hydrodynamic forces on an AUV in ocean currents, especially the drag force ($F_D$), lift force ($F_L$), and virtual mass force ($F_{VM}$),  which are displayed in Eq.~\eqref{eq3}, Eq.~\eqref{eq4}, and Eq.~\eqref{eq5}, respectively.
\begin{equation}
\label{eq3}
F_D = \frac{1}{2} \rho u^2 C_D A,
\end{equation}
where \(\rho\) represents fluid density, \(u\) is the relative velocity, \(C_D\) is the drag coefficient, and \(A\) is the frontal area.
\begin{equation}
\label{eq4}
F_L = \frac{1}{2} \rho u^2 C_L A,
\end{equation}
where \(C_L\) is the lift coefficient.
\begin{equation}
\label{eq5}
F_{VM} = \rho C_{VM} V \frac{\partial u}{\partial t},
\end{equation}
where \(C_{VM}\) is the virtual mass coefficient, and \(V\) represents the displaced volume.

These forces constitute the external force term \(\mathrm{F}\) in the Reynolds-Averaged Navier-Stokes (RANS) equation:
\begin{equation}
\label{eq6}
\rho \left( \frac{\partial \mathrm{u}}{\partial t} + \mathrm{u} \cdot \nabla \mathrm{u} \right) = -\nabla p + \mu \nabla^2 \mathrm{u} + \nabla \cdot (-\rho \overline{\mathrm{u}'\mathrm{u}'}) + \mathrm{F},
\end{equation}
where \(p\) is pressure, \(\mu\) is dynamic viscosity, and \(\overline{\mathrm{u}'\mathrm{u}'}\) represents the Reynolds stress tensor.

Eqs.~\eqref{eq4}--\eqref{eq6} collectively compute the time-varying hydrodynamic loads and their resultant force vector acting on the AUV.

To model the collisions between AUVs, we formulate a smooth repulsive force mechanism based on continuous contact mechanics. 
The smooth penetration depth function is defined as:
\begin{equation}
\label{eq_penetration}
\sigma(d) = k \cdot \log\left(1 + \exp\left(-\frac{d - (r_a + r_b)}{k}\right)\right),
\end{equation}
where $d = \|\mathrm{p}_a - \mathrm{p}_b\| \in \mathbb{R}^+$ denotes the Euclidean distance between AUV, $r_a, r_b \in \mathbb{R}^+$ represent their respective collision radii, and $k > 0$ is the smoothing parameter that governs the transition sharpness near contact boundaries. 

The collision force vector $\mathrm{F}_{\text{collision}}$ is then expressed as:
\begin{equation}
\label{eq_force}
\mathrm{F}_{\text{collision}} = F_{\text{contact}} \cdot \sigma(d) \cdot \frac{\mathrm{p}_a - \mathrm{p}_b}{\|\mathrm{p}_a - \mathrm{p}_b\|},
\end{equation}
where $F_{\text{contact}} > 0$ is the contact stiffness coefficient that controls the repulsive force strength, and $\frac{\mathrm{p}_a - \mathrm{p}_b}{\|\mathrm{p}_a - \mathrm{p}_b\|}$ is the unit direction vector from AUV $b$ to AUV $a$. 
Eqs.~\eqref{eq_penetration}--\eqref{eq_force} provide a smooth and continuous collision model that ensures stable force computations during both numerical simulation and RL training in multi-AUV systems.

\subsection{Markov Decision Process Modeling}\label{Section:3-3}

In the decision-making process for multi-AUV systems, the environment can be defined as an MDP comprising five components: state space $\mathcal{S}$, action space $\mathcal{A}$, state transition probability $\mathcal{P}$, reward function $\mathcal{R}$, and discount factor $\gamma$, as expressed in Eq.~\eqref{eq99}:
\begin{equation}
\label{eq99}
\mathcal{M} = \langle \mathcal{S}, \mathcal{A}, \mathcal{P}, \mathcal{R}, \gamma \rangle
\end{equation}

The state space $\mathcal{S}$ encodes comprehensive environmental awareness via sonar-derived spatial relationships between AUVs and surrounding objects (including obstacles and AUVs). 
Formulated as $\mathcal{S} = \{s_1, s_2, \dots, s_{|\mathcal{S}|}\}$, each state vector $s_i = (\eta_i, \phi_i, o_i), i=1,2,\cdots,|\mathcal{S}|$ integrates three orthogonal components: the ego-motion state $\eta_i = (p_{\text{pos}}, p_{\text{vel}}) \in \mathbb{R}^6$ capturing the $i$-th AUV's 3D position $p_{\text{pos}}$ and velocity $p_{\text{vel}}$, the perceptual state $\phi_i$ encoding processed environmental features from acoustic sensing, and the observation state $o_i = (\kappa_i, \sigma_i, \lambda_i) \in \mathbb{R}^{N_k \times d_k + N_\sigma \times d_\sigma + N_\lambda \times d_\lambda}$ representing relative positions of $N_k$ tracked targets, $N_\sigma$ neighboring AUVs, and $N_\lambda$ environmental landmarks.

The action space $\mathcal{A}$ defines the continuous control inputs for multi-AUV systems, formalized as $\mathcal{A} = \{a_1, a_2, \dots, a_{|A|}\}$ where each action vector $a_i = [a_x, a_y, a_z]^\top \in \mathbb{R}^3, i=1,2,\cdots,|\mathcal{A}|$ represents normalized thrust commands along the Cartesian axes. 
These control signals are generated by a policy network with $\tanh$ activation, ensuring bounded outputs within $[-1, 1]$ to maintain physical feasibility of propulsion commands while enabling precise maneuverability in complex underwater environments.

The state transition probability $\mathcal{P}(s'|s, a)$ follows physically-constrained kinematic models that govern the evolution of multi-AUV systems. 
The discount factor $\gamma \in [0, 1)$ modulates the temporal credit assignment in policy evaluation, where $\gamma$ approaching one denotes multi-AUV systems prioritize long-term cumulative rewards through strategic path planning, while $\gamma$ approximating zero emphasizes that the systems prefer the short-term reward.


In our work, the reward function $\mathcal{R}$ incorporates multiple performance factors through a weighted composite structure. 
We specifically integrate tracking accuracy metrics, collision avoidance penalties, and environmental constraint satisfaction to enhance multi-AUV coordination effectiveness. 
The proposed reward aims to balance the exploration-exploitation tradeoffs while maintaining stability during policy optimization.
The details about the reward design are showcased in Section \ref{Section:5}.


\section{Hierarchical Multi-AUV MARL Based on Supervised Diffusion}\label{Section:4}

This section introduces generative AI and proposes a hierarchical multi-AUV MARL framework based on a supervised diffusion model, aiming to mitigate training instability caused by heterogeneous sample quality in experience replay buffers. 
To address the complex coordination and learning challenges in underwater multi-agent settings, we design a structured four-layer architecture that integrates centralized guidance with decentralized execution. 
The proposed hierarchy not only stabilizes training but also enhances scalability and operational coherence between distributed AUVs.

\begin{figure*}[bth]
	\centering
	\includegraphics[width=0.9\linewidth]{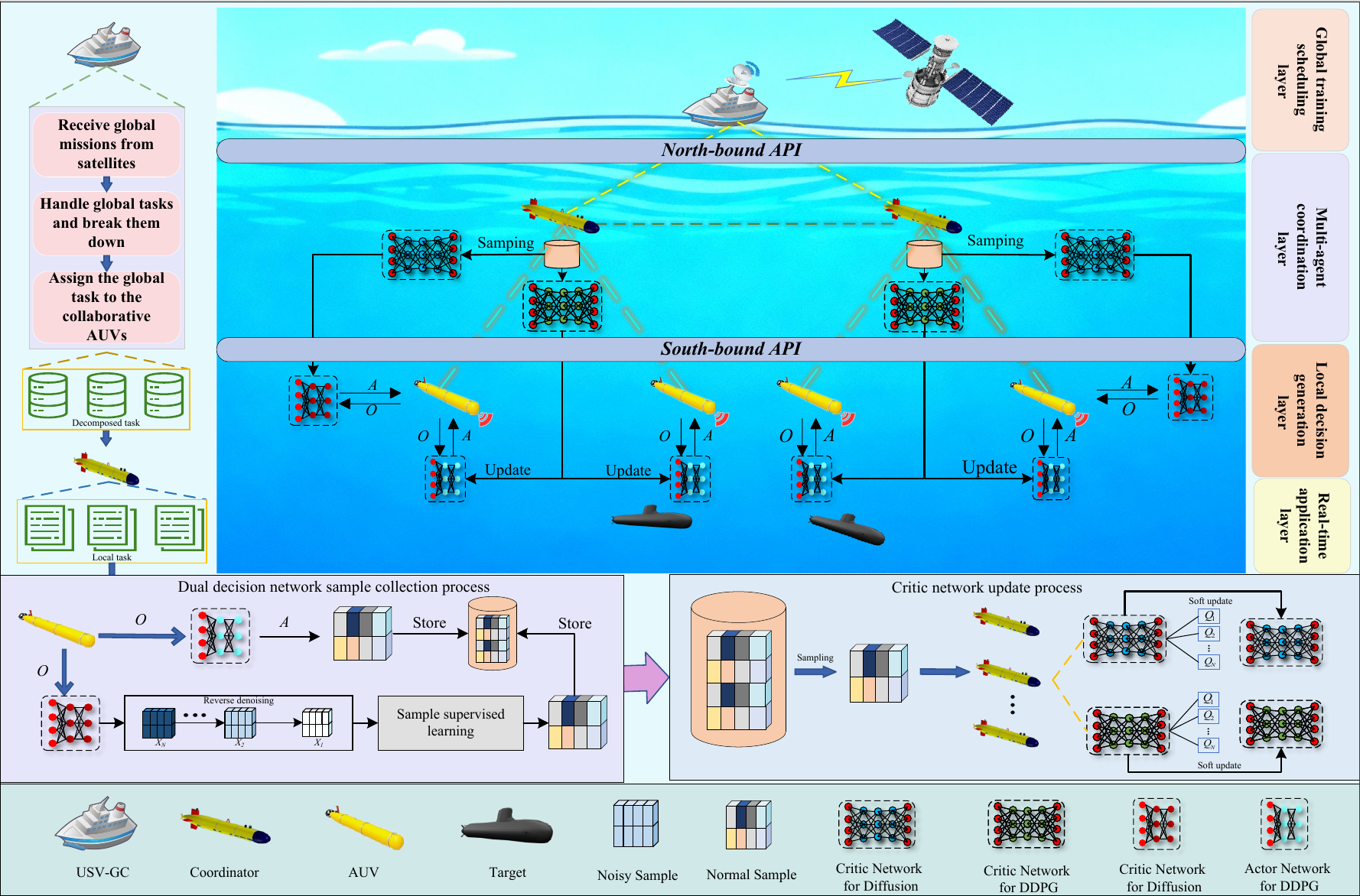}
	\caption{Architecture of hierarchical multi-AUV MARL framework based on supervised diffusion}
	\label{fig1}
\end{figure*}

\subsection{Proposed Hierarchical Multi-AUV MARL Architecture}\label{Section:4-1}

This subsection details the four-layer architecture of the proposed hierarchical framework, as illustrated in Fig. \ref{fig1}. 
The design decouples system functions into specialized layers for global training scheduling, multi-agent coordination, local policy generation, and real-time execution. Building upon the software-defined MARL concept from our prior work \cite{9226088,10621453}, this structure establishes clear functional boundaries with bidirectional information flow. 
It is engineered to enable adaptive decision-making and robust integration, directly addressing the core challenges of exploration efficiency and training stability in distributed underwater operations.

\textbf{Global Training Scheduling Layer:}
In this layer, the Unmanned Surface Vessel-based Global Controller (USV-GC) functions as the centralized mission controller (as shown in Fig. \ref{fig1}), receiving global task specifications via satellite communications. 
The USV-GC utilizes a northbound interface to decompose the global mission into spatially bounded sub-tasks according to operational regions. 
These sub-tasks are subsequently allocated to designated underwater coordinators based on their territorial assignments.
Each coordinator executes region-specific operations within its assigned underwater domain, implementing underwater acoustic communication and multi-target tracking procedures while complying with physical operation constraints.

\textbf{Multi-AUV Coordination Layer:} 
This layer processes sub-task assignments from upper tiers and executes localized information processing. 
In this layer, the coordinator nodes establish bidirectional communication with adjacent-region peers to enable cross-domain data exchange. 
Each coordinator decomposes tasks into executable primitives and assigns them to AUVs via capability-aware task assignment.
Meanwhile, this layer also implements a centralized training paradigm where coordinators integrate global evaluation networks that aggregate AUV state information for policy optimization.
Further, a unified experience replay buffer has been deployed in this layer to enhance inter-AUV knowledge transfer.
To guarantee a synchronized global view, the periodic synchronization protocol is deployed to transmit processed local data upward to global schedulers and laterally to peer coordinators.

\textbf{Local Policy Generation Layer:} 
As the policy execution unit, this layer receives distributed tasks from the coordination layer and initiates local training. 
Each AUV employs a diffusion-based generative policy that produces exploratory action distributions through multi-step denoising.
The interaction samples are stored in local experience buffers, where policy optimization combines behavior cloning and Q-learning. 
Furthermore, the DDPG networks are refined via diffusion model-guided loss to approach optimal behaviors, while bidirectional communication with upper layers maintains consistent state representations for coordinated decision-making.

\textbf{Real-Time Execution Layer:} 
This layer executes optimized policies through the AUVs.
Each AUV integrates dual decision architectures: a diffusion model and a DDPG network. 
During environmental interaction, the diffusion model first generates diverse exploration trajectories. 
These trajectories, combined with real interaction data from the DDPG, populate the shared experience replay buffer. 
The proposed dual decision architectures are jointly trained on sampled experiences, where the diffusion model provides supervisory signals for behavior cloning while the DDPG optimizes action selection.
This synergistic framework accelerates target acquisition and obstacle avoidance capabilities, yielding rapid convergence of cumulative rewards during training.
For details about our proposed generative MARL algorithm, please refer to the concept in Section \ref{Section:4-2}.

In summary, the proposed generative hierarchical multi-AUV MARL architecture implements a centralized training with a decentralized execution framework. 
The integration of diffusion models and DDPG networks as dual decision-making architectures within each AUV yields accelerated convergence stability for distributed underwater tracking tasks. 
This paradigm simultaneously enhances system cohesion while substantially reducing inter-module coupling complexity.

\subsection{Proposed Generative Multi-AUV MARL Algorithm}\label{Section:4-2}
\begin{figure}[bth]
	\centering
	\includegraphics[width=1.0\linewidth]{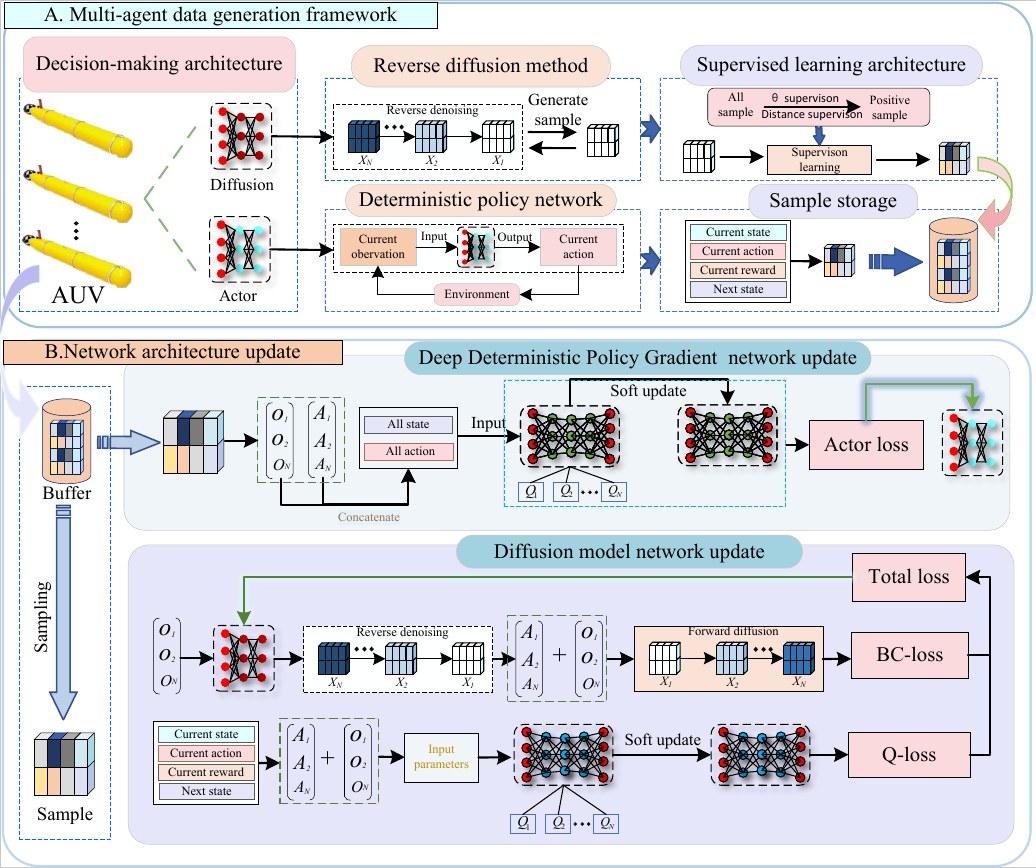}
	\caption{Workflow of proposed generative multi-AUV MARL algorithm}
	\label{fig2}
\end{figure}

To address reward oscillation and convergence challenges in underwater multi-agent training under dynamically complex environments, this paper proposes a Supervised Diffusion model-based Multi-Agent Reinforcement Learning algorithm (SDA-MARL), as shown in Fig. \ref{fig2}. 
The proposed algorithm implements a dual decision architecture combining DDPG with diffusion model.
This framework employs the diffusion component to generate high-quality synthetic experiences that enhance replay buffer diversity, while a supervised learning mechanism constrains diffusion model training using Q-value feedback to optimize sample quality. 
This integrated approach fundamentally stabilizes training dynamics by regulating the experience distribution at its source.

\subsubsection{Positive Sample Generation Based on Diffusion Model}\label{Section:4-1-1}


In conventional multi-AUV MARL frameworks, multi-AUV systems typically collect environmental interaction experiences for replay buffer storage.
However, this manner exhibits inherent limitations: immature policies during early training phases generate low-quality experiences that accumulate in the buffer and propagate through policy updates, inducing value estimation bias and policy gradient oscillations that compromise convergence stability and speed. 
To overcome these limitations, this paper introduces a collaborative diffusion model-based training architecture that systematically enhances robustness and convergence performance through structured experience generation and quality-aware filtering mechanisms.

In the proposed SDA-MARL algorithm, the experience replay buffer maintains a dual-component structure.
One partition consists of authentic interaction trajectories generated by the DDPG decision component exploring the environment.
The other partition comprises synthetic experiences produced by diffusion models.
To enable decentralized training efficiency, each AUV independently hosts a paired architecture containing both a DDPG network and a diffusion model, establishing localized training units capable of joint experience optimization without centralized coordination overhead.

In the diffusion model deployed on AUV, data generation follows a dual-phase process: forward diffusion progressively injects noise into clean samples. 
At the same time, a neural network executes reverse denoising to reconstruct data distributions. This mechanism generates novel samples consistent with the original training distribution, thereby enhancing diversity while preserving sample authenticity.

In summary, the diffusion model in the proposed SDA-MARL includes the following three steps to generate the samples:

\textbf{Step 1: Forward Diffusion in Action Space}

We define the noise schedule $\{\beta_t\}_{t=1}^T$ and the corresponding signal retention coefficients:
\begin{equation}
\alpha_t = 1 - \beta_t, 
\qquad
\bar{\alpha}_t = \prod_{s=1}^t \alpha_s,
\end{equation}
where $T$ is the total number of diffusion steps,
$t$ denotes the current timestep,
$\beta_t$ controls the injected noise at timestep $t$,
$\alpha_t$ denotes the per-step signal retention,
and $\bar{\alpha}_t$ is the cumulative signal retention from timestep $1$ to $t$.

Given an original action sample $\mathrm{x}_0$, the $t$-step forward diffusion admits the closed-form Gaussian distribution:
\begin{equation}
q(\mathrm{x}_t \mid \mathrm{x}_0)
= \mathcal{N}\!\bigl(
\mathrm{x}_t;\,
\sqrt{\bar{\alpha}_t}\,\mathrm{x}_0,\,
(1-\bar{\alpha}_t)\mathrm{I}
\bigr),
\end{equation}
where $\mathrm{x}_0$ is the original clean action,
$\mathrm{x}_t$ is the noisy action at timestep $t$,
$\mathcal{N}(\boldsymbol{\mu}, \boldsymbol{\Sigma})$ denotes a multivariate Gaussian distribution with mean $\boldsymbol{\mu}$ and covariance $\boldsymbol{\Sigma}$, and $\mathrm{I}$ is the identity matrix.

\textbf{Step 2: State-Conditioned Reverse Denoising and Sampling}

In diffusion-based reinforcement learning, the policy reconstructs actions by reversing the diffusion process in the action space.

Starting from a noisy action $\mathrm{x}_t$ and conditioning on the current state $\mathrm{s}_t$, the network $\boldsymbol{\epsilon}_\theta$ predicts the noise injected at timestep $t$ and thereby yields an estimate of the clean action:
\begin{equation}
\hat{\mathrm{x}}_0
= \frac{1}{\sqrt{\bar{\alpha}_t}}
\left(
\mathrm{x}_t
- \sqrt{1-\bar{\alpha}_t}\,
\boldsymbol{\epsilon}_\theta(\mathrm{x}_t, t, \mathrm{s}_t)
\right),
\end{equation}
where $\mathrm{s}_t$ is the current state observation,
$\boldsymbol{\epsilon}_\theta(\mathrm{x}_t, t, \mathrm{s}_t)$ is a neural network that predicts the noise component conditioned on the noisy action, timestep, and state,
and $\hat{\mathrm{x}}_0$ is the estimated clean action.

The model-predicted mean of the reverse diffusion transition is given by
\begin{equation}
\boldsymbol{\mu}_\theta(\mathrm{x}_t, t, \mathrm{s}_t)
= \frac{\sqrt{\bar{\alpha}_{t-1}}(1-\bar{\alpha}_t)}{1-\bar{\alpha}_{t-1}}\,\mathrm{x}_t
+ \frac{\sqrt{\alpha_t}\,\beta_t}{1-\bar{\alpha}_{t-1}}\,\hat{\mathrm{x}}_0,
\end{equation}
where $\bar{\alpha}_{t-1}$ is the cumulative signal retention up to timestep $t-1$,
and the coefficients are functions of the noise schedule parameters $\alpha_t$, $\bar{\alpha}_t$, and $\beta_t$.

At inference time, the policy starts from pure Gaussian noise in the action space:
\begin{equation}
\mathrm{x}_T \sim \mathcal{N}(\mathrm{0}, \mathrm{I}),
\end{equation}
where $\mathrm{x}_T$ denotes the initial noisy action at the final timestep $T$,
and $\mathrm{0}$ is the zero vector.

For each timestep $t$ from $T$ down to $1$, the policy iteratively applies state-conditioned denoising.
At each step, the reverse transition samples
\begin{equation}
\mathrm{x}_{t-1}
= \frac{1}{\sqrt{\alpha_t}}\,\mathrm{x}_t
- \frac{\beta_t}{\sqrt{1-\bar{\alpha}_t}}\,
\boldsymbol{\epsilon}_\theta(\mathrm{x}_t, t, \mathrm{s}_t)
+ \sigma_t \mathrm{z},
\end{equation}
where $\mathrm{x}_{t-1}$ is the denoised action at timestep $t-1$,
$\sigma_t$ is a time-dependent standard deviation determined by the noise schedule,
and $\mathrm{z} \sim \mathcal{N}(\mathrm{0}, \mathrm{I})$ is standard Gaussian noise sampled independently at each step.

\textbf{Step 3: Training the State-Conditioned Diffusion Policy}

To train the diffusion-based policy, the noise prediction network is optimized using transition samples collected in the replay buffer $\mathcal{B}$:
\begin{equation}
\mathcal{L}_{\text{train}}
= \mathbb{E}_{(\mathrm{s}_t, \mathrm{a}_t) \sim \mathcal{B},\, \boldsymbol{\epsilon},\, t}\left[
\left\|
\boldsymbol{\epsilon}
- \boldsymbol{\epsilon}_\theta\bigl(
\sqrt{\bar{\alpha}_t}\,\mathrm{a}_t
+ \sqrt{1-\bar{\alpha}_t}\,\boldsymbol{\epsilon},\,
t,\,
\mathrm{s}_t
\bigr)
\right\|^2
\right],
\end{equation}
where $\mathcal{B}$ is the replay buffer storing transition samples,
$\mathrm{a}_t$ is the ground-truth action sampled from the replay buffer,
$\boldsymbol{\epsilon} \sim \mathcal{N}(\mathrm{0}, \mathrm{I})$ is the noise used in the forward diffusion, and $\mathcal{L}_{\text{train}}$ is the training loss that measures the discrepancy between the predicted noise and the actual noise injected during forward diffusion.

\subsubsection{Network Update Based on Diffusion Q-Driven Behavioral Clone}\label{Section:4-1-2}
\paragraph{Q Learning-Guided Behavior Cloning Network}\label{Section:4-1-2-1}
\hfill


The diffusion actor employs a dual loss: behavior cloning minimizes action distribution divergence for stability, while Q-loss maximizes expected returns to guide the policy toward high-value regions.
Their joint optimization enables sustained performance improvement without sacrificing training stability. The total actor loss combines both objectives:

\begin{equation}
\label{eq23}
\mathcal{L}_{\text{actor}} = \mathcal{L}_{\text{BC}} + \eta \cdot \mathcal{L}_Q,
\end{equation}
where $\mathcal{L}_{\text{actor}}$ denotes the total Actor loss, $\mathcal{L}_{\text{BC}}$ is the behavior cloning loss, $\mathcal{L}_Q$ represents the Q-loss, and $\eta > 0$ is the weighting coefficient that balances the contribution of the Q-loss.

The per-sample diffusion loss measures the mean squared error between the injected noise and its prediction by the diffusion model:

\begin{equation}
\label{eq25}
\mathcal{L}_{\text{diffusion}}(a,s) = \mathbb{E}_{\substack{t \sim \text{Uniform}(0,T) \\ \boldsymbol{\epsilon} \sim \mathcal{N}(\mathrm{0},\mathrm{I})}} \left[ \left\| \boldsymbol{\epsilon} - \boldsymbol{\epsilon}_\theta(\mathrm{a}_t, t, s) \right\|_2^2 \right],
\end{equation}
where $\boldsymbol{\epsilon}$ is standard Gaussian noise, $\mathrm{a}_t = \sqrt{\bar{\alpha}_t}\,a + \sqrt{1 - \bar{\alpha}_t}\,\boldsymbol{\epsilon}$ is the noisy action at step $t$, and $\boldsymbol{\epsilon}_\theta(\cdot)$ is the noise prediction network conditioned on state $s$.

Note that $\mathcal{L}_{\text{BC}}$ is computed as the expected diffusion denoising loss over actions sampled from the replay buffer:

\begin{equation}
\label{eq24}
\mathcal{L}_{\text{BC}} = \mathbb{E}_{(s,a) \sim \mathcal{D}} \left[ \mathcal{L}_{\text{diffusion}}(a,s) \right].
\end{equation}

The Q-loss provides value-based guidance by encouraging $\pi_\theta$ to maximize the estimated Q-value, using a randomly selected Critic network from a clipped double-Q ensemble for stability:

\begin{equation}
\label{eq26}
\mathcal{L}_Q = - \mathbb{E}_{s \sim \mathcal{D}} \big[ Q_i(s, \pi_\theta(s)) \big],
\quad i \in \{1, 2\},
\end{equation}
where the index $i$ is randomly sampled from $\{1, 2\}$ at each update step to mitigate overestimation bias.

\paragraph{Dual Q-Network Architecture for Value Estimation}\label{Section:4-1-2-2}
\hfill

To mitigate Q-value overestimation, the value network employs a dual Q-network architecture. 
This architecture leverages two Critic networks operating in tandem to learn more accurate value functions. 
Specifically, each Critic network minimizes the error between its current value prediction and a target value, thereby progressively reducing estimation bias. 
The aforementioned mechanism provides Actor policy updates with stable and unbiased value gradients.

The Critic network is trained using a double-Q architecture to mitigate overestimation bias. The total Critic loss is the sum of the individual losses for the two Q-networks:

\begin{equation}
\label{eq27}
\mathcal{L}_{\text{critic}} = \mathcal{L}_{Q_1} + \mathcal{L}_{Q_2},
\end{equation}
where $\mathcal{L}_{\text{critic}}$ denotes the total Critic loss, and $\mathcal{L}_{Q_1}$, $\mathcal{L}_{Q_2}$ represent the losses for the two Q-networks.

Each Q-network is trained to minimize the mean squared error between its prediction and a target Q-value computed from the replay buffer:

\begin{equation}
\label{eq28}
\mathcal{L}_{Q_i} = \mathbb{E}_{(s,a,r,s',d) \sim \mathcal{D}} \left[ \left( Q_i(s, a) - y \right)^2 \right], \quad i \in \{1, 2\},
\end{equation}
where $Q_i(s, a)$ estimates the expected cumulative return for taking action $a$ in state $s$, and $y$ is the target value.

The target value $y$ incorporates a clipped double-Q backup to stabilize learning, using actions that maximize the target Q-functions over a candidate action set $\mathcal{A}$ (e.g., generated by the diffusion policy):

\begin{equation}
\label{eq29}
y = r + \gamma (1 - d) \cdot \min_{k \in \{1,2\}} \max_{a_j \in \mathcal{A}} Q_k^{\text{target}}(s', a_j),
\end{equation}
where \( r \) is the immediate reward, \( \gamma \in [0,1) \) is the discount factor, \( d \in \{0,1\} \) is the episode termination flag, and \( Q_k^{\text{target}} \) denotes the target networks updated via exponential moving averages.

In Eq. \eqref{eq29}, the $\min$ operator over dual Critics and $\max$ over action space $\mathcal{A}$ collectively suppresses overestimation bias while maintaining policy improvement guarantees.

Totally, Eqs. \eqref{eq27}--\eqref{eq29} formalize the value network architecture, which evaluates the long-term value of generated actions to provide critical guidance signals for policy optimization.

\paragraph{Supervised Learning-Driven Action Selection}\label{Section:4-1-2-3}
\hfill

To ensure the quality and utility of the sample produced by the diffusion model, this work further introduces a supervised learning mechanism that guides model outputs through labeled experience filtering. 
In our work, the high-quality actions identified via behavioral benchmarking are assigned positive labels and stored in the experience replay buffer.

Note that a behavioral criterion is proposed to evaluate the action quality by comparing trajectory segments. 
For an AUV transitioning from position $\mathrm{p}_{\text{prev}}$ to $\mathrm{p}_{\text{curr}}$ toward target $\mathrm{p}_{\text{target}}$, we define the \textit{displacement vector} $\Delta \mathrm{p}_i$ (actual movement) and the \textit{target vector} $\Delta \mathrm{p}_{\text{target}}$ (desired direction to the goal):

\begin{equation}
\label{eq30}
\Delta \mathrm{p}_i = \mathrm{p}_{\text{curr}} - \mathrm{p}_{\text{prev}}, \quad 
\Delta \mathrm{p}_{\text{target}} = \mathrm{p}_{\text{target}} - \mathrm{p}_{\text{prev}}.
\end{equation}

To ensure the labels, the following constraints have to be imposed:
\begin{equation}
\label{eq31}
\|\Delta \mathrm{p}_i\| > \epsilon_{\text{min}}, \quad \epsilon_{\text{min}} > 0.
\end{equation}

Here, we utilize two approaches, named displacement vectors evaluation and target approaching evaluation, to evaluate the actions.

In displacement vectors evaluation, the action quality is evaluated by the cosine similarity $\cos(\cdot)$ between displacement vectors $\Delta \mathrm{p}_i$ and $\Delta \mathrm{p}_{\text{target}}$:
\begin{equation}
\label{eq32}
\cos(\theta) = \frac{\Delta \mathrm{p}_i \cdot \Delta \mathrm{p}_{\text{target}}}{\|\Delta \mathrm{p}_i\| \cdot \|\Delta \mathrm{p}_{\text{target}}\|}.
\end{equation}
Obviously, the larger the value of Eq. \eqref{eq32}, the larger score the action is achieved.

For target approaching evaluation, an action is demonstrated to be approaching the target only when the following $converge$ is equal to 1:
\begin{equation}
\label{eq33}
\mathit{converge} =
\begin{cases}
1, & \text{if } \|\mathrm{p}_{\text{curr}} - \mathrm{p}_{\text{target}}\| < \|\mathrm{p}_{\text{prev}} - \mathrm{p}_{\text{target}}\|, \\
0, & \text{otherwise.}
\end{cases}.
\end{equation}

Totally, an action is demonstrated to be available only when the following $valid$ is equal to 1:
\begin{equation}
\label{eq34b}
\mathit{valid} =
\begin{cases}
1, & \text{if } \mathit{aligned}=1 \ \land\ \mathit{converge}=1,\\
0, & \text{otherwise,}
\end{cases},
\end{equation}
where $aligned$ is the following:
\begin{equation}
\label{eq34a}
\mathit{aligned} =
\begin{cases}
1, & \text{if } \cos(\theta) > \cos(\theta_{\text{threshold}}),\\
0, & \text{otherwise,}
\end{cases},
\end{equation}
where $\theta_{\text{threshold}} \in (0, \pi/2)$ is a predefined angular tolerance. 
Only experiences satisfying Eq. \eqref{eq34b} receive positive labels and populate the replay buffer, ensuring high-quality training data while maintaining distributional diversity.

\begin{algorithm}
\caption{Tracking Quality Assessment}
\label{alg:quality_assessment}
\begin{algorithmic}[1]
\Require Number of agents $N$, directional tolerance $\theta_{\text{tol}}$, displacement threshold $\epsilon_{\text{min}}$, position vectors $\{\mathrm{p}_{\text{prev}}^{(i)}, \mathrm{p}_{\text{curr}}^{(i)}\}$
\Ensure Quality labels $\mathcal{L} = [\mathcal{L}_1, \mathcal{L}_2, \dots, \mathcal{L}_N]$
\State Initialize directional tolerance $\theta_{\text{tol}}$ and displacement threshold $\epsilon_{\text{min}}$ by Eq. \eqref{eq31}.
\For{$i = 1$ to $N$}
    \State Compute displacement vector $\Delta \mathrm{p}_i$ by Eq. \eqref{eq30}.
    \If{$\|\Delta \mathrm{p}_i\| < \epsilon_{\text{min}}$}
        \State Assign $\mathcal{L}_i = 0$.
        \State \textbf{continue}.
    \EndIf
    \State Calculate cosine similarity by Eq. \eqref{eq32}.
    \State Evaluate target approach by Eq. \eqref{eq33}.
    \State Determine alignment condition by Eq. \eqref{eq34a}.
    \State Set $\mathcal{L}_i = 1$ if conditions in Eq. \eqref{eq34b} are satisfied.
\EndFor
\State \Return quality labels $\mathcal{L}$.
\end{algorithmic}
\end{algorithm}

\begin{algorithm}
\caption{High-Value Experience Harvesting}
\label{alg:experience_harvesting}
\begin{algorithmic}[1]
\Require Number of episodes $n_{\text{ep}}$, episode horizon $H$, exploration noise $\sigma$, minimum valid agent ratio $\alpha = 1/3$
\Ensure Total harvested samples $N_{\text{samples}}$
\State Initialize episode horizon $H$ and exploration noise $\sigma$.
\For{$e = 1$ to $n_{\text{ep}}$}
    \State Reset environment state $\mathrm{s}$.
    \For{$t = 1$ to $H$}
        \State Obtain agent positions $\{\mathrm{p}_{\text{prev}}^{(i)}\}$.
        \State Generate action $\mathrm{a} = \pi_\theta(\mathrm{s}) + \mathcal{N}(0,\sigma)$.
        \State Execute action and obtain new state $\mathrm{s}'$.
        \State Record new positions $\{\mathrm{p}_{\text{curr}}^{(i)}\}$.
        \State Evaluate quality by Algorithm~\ref{alg:quality_assessment} with inputs $(N, \theta_{\text{tol}}, \epsilon_{\text{min}}, \{\mathrm{p}_{\text{prev}}^{(i)}\}, \{\mathrm{p}_{\text{curr}}^{(i)}\})$.
        \If{at least $\lfloor N/3 \rfloor$ agents have $\mathcal{L}_i = 1$ (valid actions per Eq.~\eqref{eq34b})}
            \State Store experience in replay buffer $\mathcal{D}$.
            \State Increment sample count $N_{\text{samples}}$.
        \EndIf
        \State Update state $\mathrm{s} \gets \mathrm{s}'$.
        \If{episode terminated}
            \State \textbf{break}.
        \EndIf
    \EndFor
\EndFor
\State \Return total harvested samples $N_{\text{samples}}$.
\end{algorithmic}
\end{algorithm}

In summary, the SDA-MARL algorithm utilizes algorithm~\ref{alg:quality_assessment} and algorithm~\ref{alg:experience_harvesting} to assess agent behavior quality and collect high-quality samples. Specifically, each agent receives a binary quality label, and a state-action pair is collected for supervision only when positive tracking behavior is observed in at least one-third of the agents.

Algorithm~\ref{alg:quality_assessment} implements tracking quality assessment by first initializing the directional tolerance $\theta_{\text{tol}}$ and displacement threshold $\epsilon_{\text{min}}$ according to Eq.~\eqref{eq31} (Line~1). For each AUV $i$, it computes the displacement vector $\Delta \mathrm{p}_i$ via Eq.~\eqref{eq30} (Line~3), and immediately discards insignificant movements with $\|\Delta \mathrm{p}_i\| < \epsilon_{\text{min}}$ (Lines~4--6). Valid trajectories are then evaluated through cosine similarity (Eq.~\eqref{eq32}, Line~8), target approach status (Eq.~\eqref{eq33}, Line~9), and directional alignment (Eq.~\eqref{eq34a}, Line~10), with the final quality label $\mathcal{L}_i$ assigned based on the conditions in Eq.~\eqref{eq34b} (Line~11).

Algorithm~\ref{alg:experience_harvesting} implements the high-value experience collection framework by initializing episode horizon $H$ and exploration noise scale $\sigma$ (Line 1). It iterates through training episodes (Line 2) and executes environmental interactions where actions are generated through policy network $\pi_\theta$ with added exploration noise (Line 6). 
The algorithm~\ref{alg:experience_harvesting} records positional data before action execution (Line 5) and after execution (Line 8), evaluates tracking quality through algorithm~\ref{alg:quality_assessment} (Line 9), and selectively stores experiences when sufficient AUVs ($\geq \lfloor N/3 \rfloor$) exhibit valid behaviors per Eq.~\eqref{eq34b} (Lines 10--12). 
This selective storage mechanism ensures the replay buffer $\mathcal{D}$ maintains high-quality training samples while preserving distributional diversity. 
The algorithm concludes by returning the total count of harvested samples $N_{\text{samples}}$ that satisfy our behavioral criteria (Line 16), thereby establishing an effective supervised learning pipeline for diffusion model training.

\section{Proposed Multi-AUV Cooperative Target Tracking Algorithm}\label{Section:5}
In this section, we present the proposed multi-AUV cooperative target tracking algorithm. 
We first elucidate the core cooperative tracking strategy that enables coordinated observation of dynamic targets through spatially distributed sensing. 
Subsequently, we comprehensively detail the complete operational workflow of the proposed multi-AUV cooperative tracking algorithm.

\subsection{Proposed Reward Function for Multi-AUV Tracking}\label{Section:5-1}
This section details the reward function design for underwater intelligent tracking within the SDA-MARL framework. 
We model the tracking problem as an MDP, where the optimization objective is transformed into maximizing the expected cumulative reward. 
To achieve robust tracking in complex underwater environments, we establish three essential reward components that collectively optimize policy learning while maintaining long-term tracking stability through adaptive environmental interactions. 
The proposed reward function is displayed as follows:

\begin{equation}
\label{eq_total}
R_i = \alpha \cdot r_{\text{pos}} + \beta \cdot r_{\text{col}} + \gamma \cdot r_{\text{land}},
\end{equation}
where $R_i$ denotes the weighted composite reward for the $i$-th AUV, with $\alpha, \beta, \gamma \in \mathbb{R}^+$ serving as independent scaling coefficients that balance positional fidelity, collision avoidance, and environmental navigation priorities. 

The fundamental distance metrics are defined as follows:~$\mathrm{p}_{\text{target}}$ denotes the target position, $\mathrm{p}_j$ represents the position of the $j$-th AUV, and $\mathrm{o}_k$ indicates the position of the $k$-th obstacle in the environment:

\begin{equation}
\label{eq_distances}
\left\{
\begin{aligned}
d_t &= \|\mathrm{p}_i - \mathrm{p}_{\text{target}}\|_2, \\
d_a &= \min_{j \neq i} \|\mathrm{p}_i - \mathrm{p}_j\|_2, \\
d_l &= \min_{k \in \mathcal{O}} \|\mathrm{p}_i - \mathrm{o}_k\|_2.
\end{aligned},
\right.
\end{equation}
where $d_t$ represents the Euclidean distance to the target, $d_a$ denotes the minimum pairwise distance to neighboring AUVs, $d_l$ is the distance to the nearest environmental obstacle, $\mathcal{O}$ is the set of obstacles with positions $\mathrm{o}_k$, and all distances are computed in the configuration space $\mathbb{R}^3$.

To maintain stable pursuit trajectories while respecting safety constraints, the positional reward component employs a dual-regime penalty structure:
\begin{equation}
\label{eq_pos}
r_{\text{pos}} = 
\begin{cases} 
w \cdot d_t - (w+1) \cdot \delta_t^{\min}, & \text{if } d_t \leq \delta_t^{\min}, \\
-d_t, & \text{otherwise}.
\end{cases},
\end{equation}
where $\delta_t^{\min} > 0$ defines the minimum safety margin, and $w > 1$ modulates the penalty intensity within the critical proximity zone.

Complementing positional control, inter-AUV collision avoidance establishes dynamic repulsive fields between cooperative AUVs:
\begin{equation}
\label{eq_col}
r_{\text{col}} = 
\begin{cases} 
d_a - \delta_a^{\min}, & \text{if } d_a < \delta_a^{\min}, \\
-d_a, & \text{otherwise}.
\end{cases},
\end{equation}
where $\delta_a^{\min} > 0$ specifies the minimum inter-AUV safety distance threshold.

Environmental obstacle avoidance is enforced through a binary penalty mechanism:
\begin{equation}
\label{eq_land}
r_{\text{land}} = 
\begin{cases}
-P_{\text{land}}, & \text{if } d_l < \delta_l^{\min}, \\
0, & \text{otherwise}.
\end{cases},
\end{equation}
where $P_{\text{land}} > 0$ is a constant penalty magnitude, $d_l$ denotes the distance between the $i$-th AUV and the nearest environmental obstacle, and $\delta_l^{\min}$ is the minimum safe navigation distance threshold.

This composite reward structure integrates the three critical components for underwater multi-AUV operations: target pursuit fidelity, inter-AUV collision avoidance, and environmental obstacle navigation. 
The synergistic combination of these terms enables stable policy learning in complex hydrodynamic environments while maintaining physically plausible behavior consistent with real-world underwater navigation constraints.

\subsection{Proposed Tracking Algorithm Based on SDA-MARL}\label{Section:5-2}
\begin{algorithm}
\caption{Proposed Multi-AUV Target Tracking Algorithm Based on SDA-MARL}
\label{alg:sdamarl_tracking}
\begin{algorithmic}[1]
\Require Number of episodes $N_{\text{ep}}$, episode length $L_{\text{ep}}$, batch size $B$, update interval $I_{\text{update}}$, minimal buffer size $S_{\min}$, soft update coefficient $\tau$, EMA update interval $I_{\text{ema}}$.
\Ensure Tracking policy based on the trained SDA-MARL.
\State Initialize environment $\mathit{env}$ and the MADDPG model $\mathit{maddpg}$.
\State Initialize replay buffer $\text{buffer}$.
\State $t_{\text{total}} \gets 0$.
\For{$e = 1$ \textbf{to} $N_{\text{ep}}$}
    \State Reset environment.
    \State Collect high-quality samples using Algorithm~\ref{alg:experience_harvesting}.
    \For{$t = 1$ \textbf{to} $L_{\text{ep}}$}
        \State Each AUV $i$ takes action $\mathrm{a}_i = \pi_{\theta_i}(\mathrm{o}_i) + \mathcal{N}_t$ based on policy and exploration.
        \State Update environment, obtain next state $\mathrm{s}'$ and reward $\mathrm{r}$.
        \State Add transition $(\mathrm{s}, \mathrm{a}, \mathrm{r}, \mathrm{s}', \text{done})$ to replay buffer with source identifier 0.
        \State Update current state $\mathrm{s} \gets \mathrm{s}'$ and increment timestep counter $t_{\text{total}} \gets t_{\text{total}} + 1$.
        \If{$\text{buffer}.\text{size}() \geq S_{\min}$ \textbf{and} $t_{\text{total}} \bmod I_{\text{update}} = 0$}
            \State Sample batch $\mathcal{B}$ from $\text{buffer}$.
            \For{each AUV $i = 1$ \textbf{to} $N$}
                \State Update normal AUV $i$'s Critic and Actor using transitions from $\mathcal{B}$ with source $\in \{0, 1\}$ (Eqs.~\eqref{eq27}, \eqref{eq29}).
                \If{transitions with source $= 1$ exist in $\mathcal{B}$}
                    \State Update diffusion AUV $i$'s Critic and Actor using transitions with source $= 1$ (Eqs.~\eqref{eq23}--\eqref{eq29}).
                    \If{diffusion agent $i$'s step counter $\bmod I_{\text{ema}} = 0$}
                        \State Update EMA model for diffusion agent $i$.
                    \EndIf
                    \State Soft update Critic target for diffusion agent $i$ by $\theta_{\text{critic\_target}} \gets (1-\tau) \theta_{\text{critic\_target}} + \tau \theta_{\text{critic}}$.
                \EndIf
            \EndFor
            \State Soft update Actor and Critic targets for all normal agents by $\theta_{\text{target}} \gets (1-\tau) \theta_{\text{target}} + \tau \theta$.
        \EndIf
        \If{done}
            \State \textbf{break}.
        \EndIf
    \EndFor
\EndFor 
\end{algorithmic}
\end{algorithm}

\begin{figure*}[bth]
	\centering
	\includegraphics[width=0.9\linewidth]{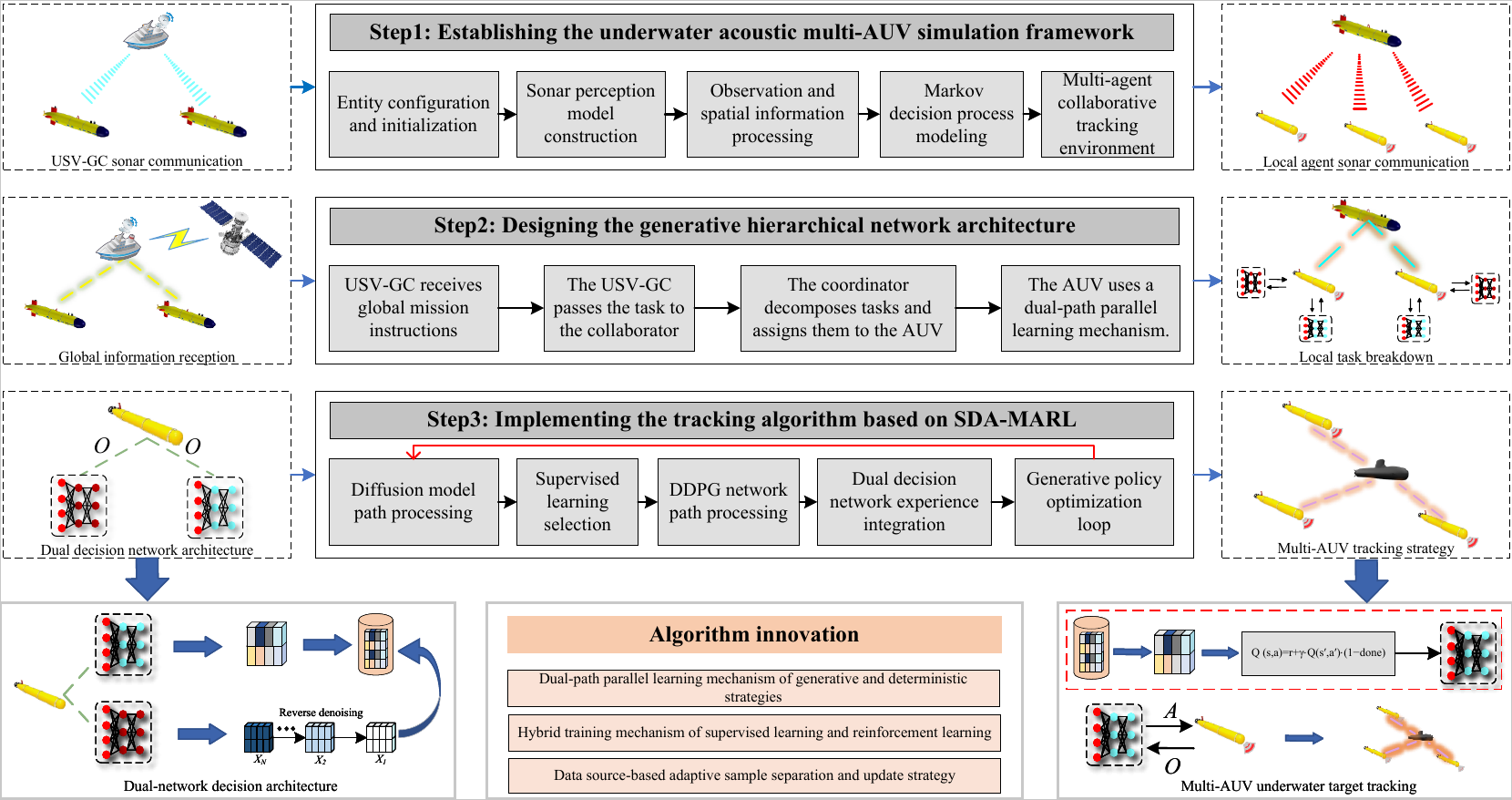}
	\caption{Proposed tracking architecture based on SDA-MARL}
	\label{fig3}
\end{figure*}

The proposed multi-AUV tracking process based on SDA-MARL can be systematically decomposed into the following three steps, as illustrated in Fig.~\ref{fig3}.

\textbf{Step 1: Establishing the underwater acoustic multi-AUV simulation framework.} 
It establishes an integrated underwater acoustic tracking framework featuring entity initialization with environmental constraints, sonar perception modeling under complex channel dynamics, spatial feature extraction of bearing/range/velocity, Markovian state-observation dynamics ($P(s_{t+1}|s_t,a_t)$, $O(o_t|s_t,a_t)$), and unified environment integration for policy training and evaluation.

\textbf{Step 2: Designing the generative hierarchical network architecture.} 
It introduces a four-layer hierarchical architecture to translate global missions into distributed AUV execution. It establishes a structured pipeline: the global layer assigns tasks, the coordination layer decomposes them, and the AUV layers implement a dual-path learning module. This design formally defines the interfaces and data flows between strategic generation and deterministic execution, providing the essential framework for implementing the SDA-MARL tracking algorithm in Step 3.

\textbf{Step 3: Implementing the tracking algorithm based on SDA-MARL.} 
At the AUV decision,i.e., local policy generation layer and real-time execution layer, stochastic exploration through diffusion-based sampling and deterministic exploitation via direct action output are adaptively fused using historical experience, with key innovations in parallel generative-deterministic learning, hybrid supervised-reinforcement training, and data source-guided adaptive sample management collectively enhancing tracking precision in complex underwater environments.

In summary, the proposed tracking algorithm is implemented in algorithm~\ref{alg:sdamarl_tracking}, which integrates all three steps within a unified multi-AUV training loop.
Algorithm~\ref{alg:sdamarl_tracking} initializes the environment, MADDPG \cite{10.5555/3295222.3295385} model, replay buffer, and timestep counter $t_{\text{total}}$ (Lines 1--3), then iterates over episodes (Line 4).
At each episode start, the environment is reset (Line 5) and algorithm~\ref{alg:experience_harvesting} harvests high-quality samples (Line 6).
Within each episode, actions are generated by the MADDPG policy with exploration noise (Line 8), transitions (s, a, r, s', done) are recorded and stored with source identifier 0 (Line 10), and the state and timestep counter are updated (Line 11).
When the buffer size exceeds $S_{\min}$ and $t_{\text{total}} \bmod I_{\text{update}} = 0$ (Line 12), a batch is sampled (Line 13) to update all agents.
Normal agents' Critics and Actors are updated via Eqs.~\eqref{eq27} and \eqref{eq29} using transitions with source identifiers 0 or 1 (Line 15).
When high-quality samples (source identifier 1) are available, diffusion agents' networks are updated using Eqs.~\eqref{eq23}--\eqref{eq29} (Line 17), with periodic Exponential Moving Average (EMA) model updates (every $I_{\text{ema}}$ steps) and Critic target soft-updates (Lines 18--20).
Each update cycle concludes by soft-updating all normal AUVs' target networks (Line 23).
This dual-update mechanism synergistically integrates conventional MADDPG training with diffusion-enhanced policy optimization to maximize tracking robustness while maintaining sample efficiency.


\section{Evaluations}\label{Section:6}
This section conducts some evaluations to demonstrate the efficiency of our proposed tracking algorithm in terms of average reward values, algorithmic effectiveness, target-tracking accuracy, etc., by comparing our proposed algorithms with some of the state-of-the-art MARL-based tracking algorithms.

\subsection{Simulation Setup}\label{Section:6-1}

All experiments are conducted on a computational platform equipped with an AMD Ryzen 9 8940HX processor, RTX 5060 GPU, and 16 GB RAM, where all the code is implemented in Python 3.10.
In the evaluations, the target moves at a predefined constant velocity, while AUVs are initially distributed in a circular ring-shaped region approximately 2 km from the target. 

To comprehensively evaluate algorithm performance, four distinct tracking scenarios are employed in the assessment: 2 AUVs tracking 1 target, 4 AUVs tracking 2 targets, 6 AUVs tracking 2 targets, and 8 AUVs tracking 3 targets. All experiments are performed within a stable underwater simulation environment.

\begin{figure*}[t!]
    \centering
    \subfloat[Scenario of 2 AUVs tracking 1 target]
    {\includegraphics[width=0.47\textwidth]{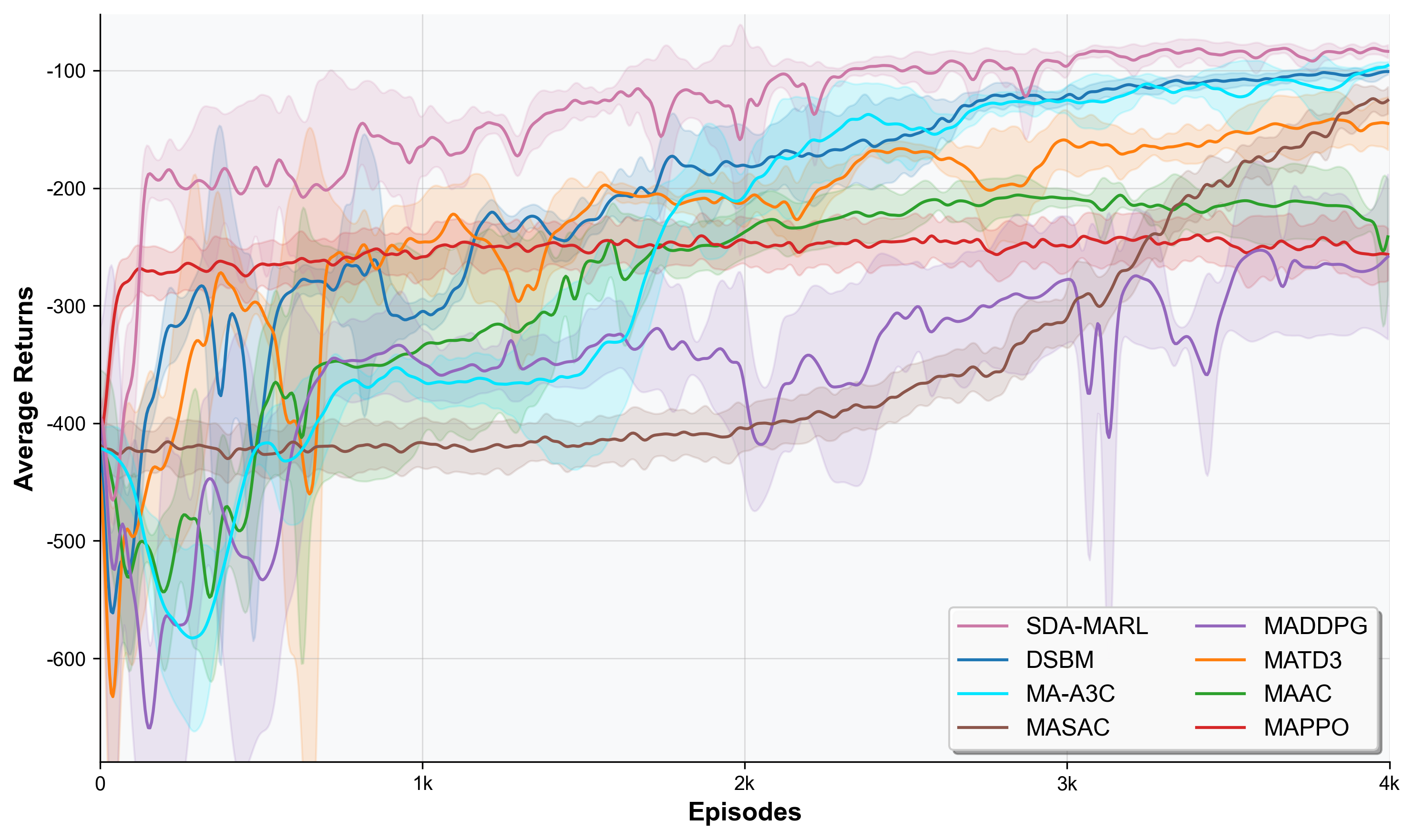}}\hfill
    \subfloat[Scenario of 4 AUVs tracking 2 targets]
    {\includegraphics[width=0.47\textwidth]{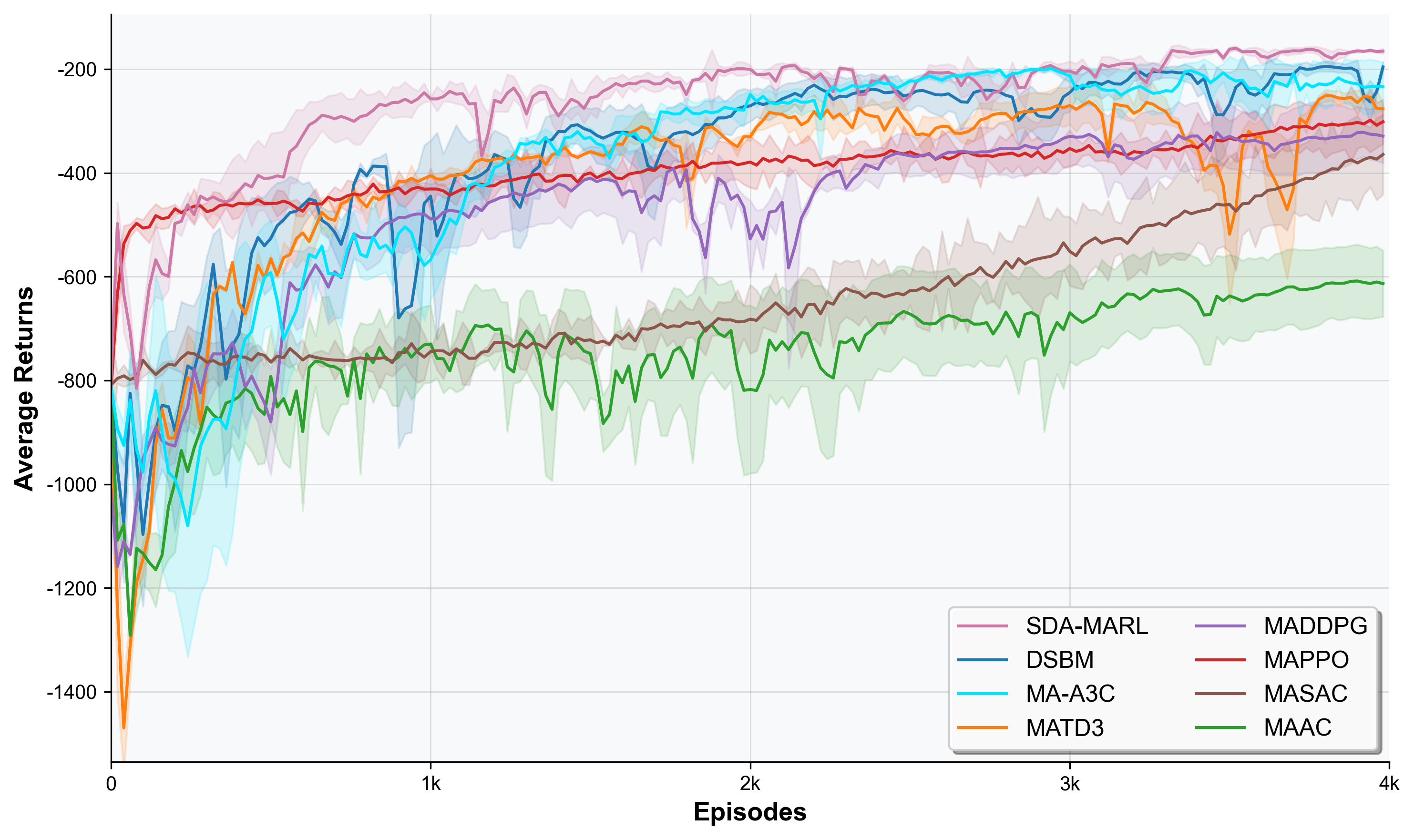}}
    
    \vspace{0.5em} 
    
    \subfloat[Scenario of 6 AUVs tracking 2 targets]
    {\includegraphics[width=0.47\textwidth]{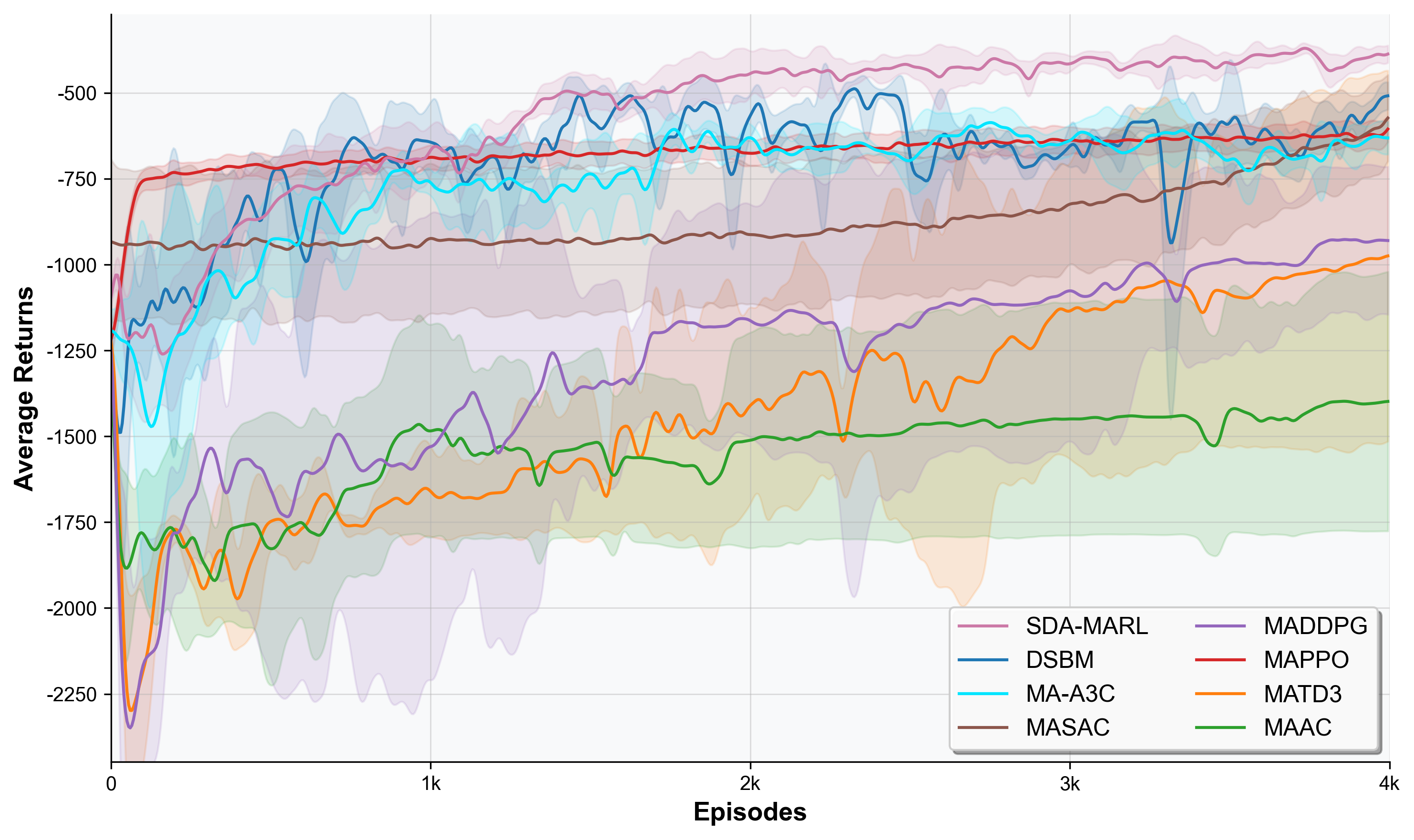}}\hfill
    \subfloat[Scenario of 8 AUVs tracking 3 targets]
    {\includegraphics[width=0.47\textwidth]{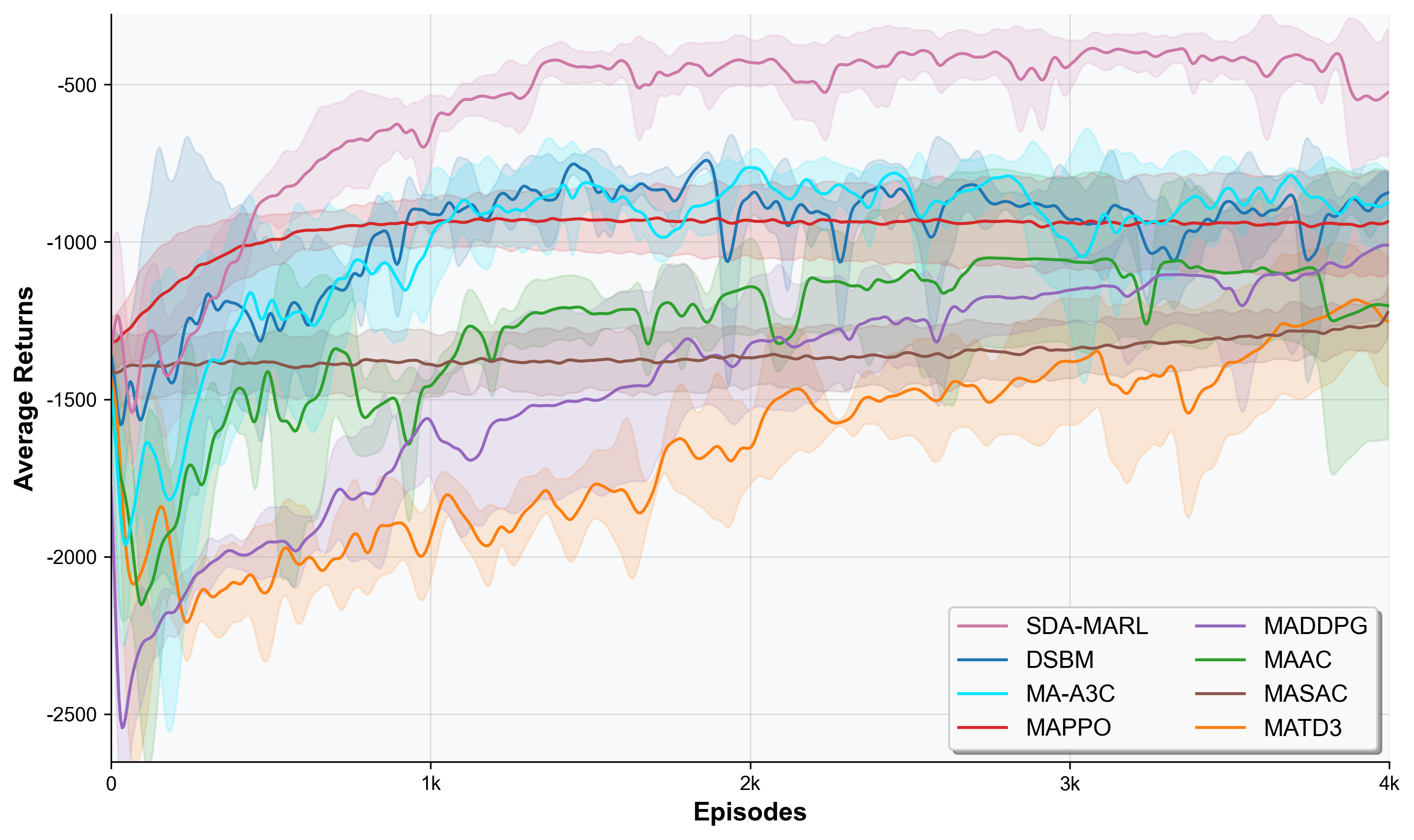}}
    
    \caption{Convergence speed evaluation}
    \label{fig6}
\end{figure*}

All the parameters in the evaluations are detailed in Table \ref{table1}.

\subsection{Results and Discussion}\label{Section:6-2}

To evaluate algorithmic effectiveness, comprehensive comparisons and analyses are conducted against state-of-the-art MARLs, including DSBM \cite{9226088}, MA-A3C \cite{10621453}, MASAC \cite{9446746}, MAPPO \cite{NEURIPS2022_9c1535a0}, MAAC \cite{doi:10.1142/S0218001422520140}, MATD3 \cite{ackermann2019reducingoverestimationbiasmultiagent}, and MADDPG.
Our approach is evaluated mainly from the following aspects: 1) convergence speed; 2) tracking accuracy; 3) velocity difference stability among multi-AUV systems; 4) path length relative to optimal trajectories; 5) diffusion steps required for strategy generation in our framework; 6) ablation studies validating each component's contribution to system performance, and 7) system availability.

\begin{table}[H]
    \centering
    \caption{Simulation parameters}
    \label{table1}
    \begin{tabular}{ccc}
    \hline
    \textbf{Parameter} & \textbf{Description} & \textbf{Value} \\ \hline
    $N_A$ & Number of AUVs & [2,4,6,8] \\
    $N_T$ & Number of Targets & [1,2,3] \\
    $l_r$ & Learning rate & $1e^{-3}$ \\
    $N_E$ & Training rounds & 4000 \\
    $N_h$ & Hidden layer neurons & 256 \\
    $\gamma$ & Discount factor & 0.95 \\
    $\tau$ & Network update coefficient & $1e^{-2}$ \\
    $d_{\min}^{\phi}$ & Minimum tracking distance & 80 m \\
    $d_{\min}^{\kappa}$ & Minimum AUV distance & 80 m \\
    $L_e$ & Episode length & 400 \\
    $B_s$ & Buffer size & 1,000,000 \\
    $U_i$ & Update interval & 400 \\
    $M_s$ & Minimal buffer size & 4000 \\
    $B$ & Batch size & 256 \\
    $\rho$ & Fluid density & 1000 kg/m$^3$ \\
    $\mu$ & Fluid viscosity & $10^{-3}$ Pa$\cdot$s \\
    $D$ & Damping factor & 0.25 \\
    $\Delta t$ & Simulation time step & 0.1 s \\
    \hline
\end{tabular}
\end{table}

\begin{table*}[t]
\centering
\caption{Tracking accuracy comparison}
\label{tab:tracking_accuracy}
\resizebox{\textwidth}{!}{%
\begin{tabular}{lcccc}
\hline
\multirow{2}{*}{\textbf{Algorithms}} & \textbf{2 AUVs Tracking 1 Target} & \textbf{4 AUVs Tracking 2 Targets} & \textbf{6 AUVs Tracking 2 Targets} & \textbf{8 AUVs Tracking 3 Targets} \\
\cline{2-5}
 & $Mean \pm SD$ & $Mean \pm SD $ & $Mean \pm SD $ & $Mean \pm SD$ \\
\hline
SDA-MARL & 66.44\%$\pm$0.06\% & 68.56\%$\pm$1.56\% & 61.38\%$\pm$0.91\% & 61.67\%$\pm$2.34\% \\
DBSM & 45.70\%$\pm$1.75\% & 47.12\%$\pm$2.82\% & 14.93\%$\pm$1.41\% & 12.69\%$\pm$1.77\% \\
MA-A3C & 41.04\%$\pm$3.58\% & 21.74\%$\pm$2.56\% & 12.53\%$\pm$1.32\% & 10.30\%$\pm$0.61\% \\
MAPPO & 12.37\%$\pm$4.88\% & 8.17\%$\pm$3.90\% & 10.50\%$\pm$1.55\% & 28.20\%$\pm$2.16\% \\
MASAC & 9.81\%$\pm$0.19\% & 5.44\%$\pm$0.12\% & 14.50\%$\pm$1.91\% & 9.04\%$\pm$0.08\% \\
MAAC & 34.37\%$\pm$1.43\% & 31.12\%$\pm$0.01\% & 16.92\%$\pm$0.13\% & 14.52\%$\pm$0.48\% \\
MATD3 & 45.10\%$\pm$0.05\% & 25.38\%$\pm$0.62\% & 19.24\%$\pm$0.02\% & 14.68\%$\pm$0.04\% \\
MADDPG & 20.31\%$\pm$0.34\% & 15.06\%$\pm$0.74\% & 9.89\%$\pm$0.35\% & 12.45\%$\pm$0.23\% \\
\hline
\end{tabular}
}
\end{table*}

\begin{table*}[t]
\centering
\caption{Velocity difference mean value comparison}
\label{tab:velocity_mean}
\resizebox{\textwidth}{!}{%
\begin{tabular}{lcccc}
\hline
\multirow{2}{*}{\textbf{Algorithms}} & \textbf{2 AUVs Tracking 1 Target} & \textbf{4 AUVs Tracking 2 Targets} & \textbf{6 AUVs Tracking 2 Targets} & \textbf{8 AUVs Tracking 3 Targets} \\
\cline{2-5}
 & $Mean$ & $Mean$ & $Mean$ & $Mean$ \\
\hline
SDA-MARL & 0.015477 & 0.017505 &  0.018757 & 0.016895 \\
DBSM & 0.030882 & 0.030721 & 0.026736 & 0.026041 \\
MA-A3C & 0.027531 & 0.030716 & 0.028926 & 0.027545 \\
MAPPO & 0.039937 & 0.035275 & 0.035289 & 0.040667 \\
MASAC & 0.038340 & 0.033087 & 0.034900 & 0.034166 \\
MAAC & 0.032703 & 0.025876 & 0.031128 & 0.035962 \\
MATD3 & 0.022570 & 0.028931 & 0.036616 & 0.043496 \\
MADDPG & 0.041735 & 0.034485 & 0.046056 & 0.044488 \\
\hline
\end{tabular}
}
\end{table*}

\begin{table*}[t]
\centering
\caption{Velocity difference standard deviation comparison}
\label{tab:velocity_std}
\resizebox{\textwidth}{!}{%
\begin{tabular}{lcccc}
\hline
\multirow{2}{*}{\textbf{Algorithms}} & \textbf{2 AUVs Tracking 1 Target} & \textbf{4 AUVs Tracking 2 Targets} & \textbf{6 AUVs Tracking 2 Targets} & \textbf{8 AUVs Tracking 3 Targets} \\
\cline{2-5}
 & $SD$ & $SD$ & $SD$ & $SD$ \\
\hline
SDA-MARL & 0.000271 & 0.000165 & 0.000062 & 0.000085 \\
DBSM & 0.000231 & 0.000183 & 0.000157 & 0.001350 \\
MA-A3C & 0.000441 & 0.000385 & 0.000145 & 0.000200 \\
MAPPO & 0.000593 & 0.000367 & 0.000565 & 0.000223 \\
MASAC & 0.000201 & 0.000075 & 0.000065 & 0.000143 \\
MAAC & 0.000039 & 0.000047 & 0.000007 & 0.000170 \\
MATD3 & 0.000133 & 0.000010 & 0.000006 & 0.000015 \\
MADDPG & 0.000080 & 0.000025 & 0.000067 & 0.000093 \\
\hline
\end{tabular}
}
\end{table*}

\begin{table*}[t]
\centering
\caption{Path length comparison}
\label{tab:path_length}
\resizebox{\textwidth}{!}{%
\begin{tabular}{lcccc}
\hline
\multirow{2}{*}{\textbf{Algorithms}} & \textbf{2 AUVs Tracking 1 Target} & \textbf{4 AUVs Tracking 2 Targets} & \textbf{6 AUVs Tracking 2 Targets} & \textbf{8 AUVs Tracking 3 Targets} \\
\cline{2-5}
 & $Mean \pm SD$ & $Mean \pm SD$ & $Mean \pm SD$ & $Mean \pm SD$ \\
\hline
SDA-MARL & $1.93 \pm 0.0053$ & $3.48 \pm 0.0044$ & $6.10 \pm 0.0341$ & $6.80 \pm 0.0257$ \\
DBSM & $2.54 \pm 0.0192$ & $4.91 \pm 0.0424$ & $6.26 \pm 0.0475$ & $6.98 \pm 0.4590$ \\
MA-A3C & $2.22 \pm 0.0385$ & $4.87 \pm 0.0641$ & $6.78 \pm 0.0405$ & $8.36 \pm 0.0686$ \\
MAPPO & $3.34 \pm 0.0330$ & $5.68 \pm 0.0459$ & $8.55 \pm 0.1274$ & $13.09 \pm 0.0997$ \\
MASAC & $3.33 \pm 0.0127$ & $4.39 \pm 0.0311$ & $10.80 \pm 0.0768$ & $11.60 \pm 0.0528$ \\
MAAC & $2.62 \pm 0.0020$ & $4.48 \pm 0.0054$ & $7.20 \pm 0.0024$ & $11.20 \pm 0.0496$ \\
MATD3 & $2.20 \pm 0.0228$ & $4.66 \pm 0.0019$ & $8.52 \pm 0.0011$ & $14.33 \pm 0.0069$ \\
MADDPG & $3.38 \pm 0.0031$ & $5.56 \pm 0.0020$ & $11.37 \pm 0.0286$ & $14.74 \pm 0.0252$ \\
\hline
\end{tabular}
}
\end{table*}

\subsubsection{\textbf{System convergence speed}}

To evaluate the convergence speed of the proposed SDA-MARL algorithm, multiple tracking experiments are conducted across various scenarios, with results presented in Figs. \ref{fig6}(a), \ref{fig6}(b), \ref{fig6}(c), \ref{fig6}(d). Overall, the proposed SDA-MARL algorithm demonstrates significantly faster convergence compared to the other seven algorithms across multiple scenarios.

Regarding convergence speed, DSBM and MA-A3C exhibit stable performance, while MAPPO also demonstrates excellent results primarily due to its centralized training architecture combined with Generalized Advantage Estimation (GAE), along with a stochastic policy maintained in its Actor network. 
This integration effectively balances global coordination and policy diversity in multi-AUV tracking tasks.

Notably, our proposed SDA-MARL manifests rapid convergence characteristics during early training stages. 
This advantage stems from the diffusion model-based behavioral cloning loss and supervised learning mechanism integrated within the framework. 
Specifically, the algorithm continuously generates high-quality action samples through the behavioral cloning loss of the diffusion model, which are systematically stored in the replay buffer for subsequent model training. 
During DDPG network updates, this behavioral cloning loss serves as a regularization term, guiding the DDPG toward the optimized strategy distribution represented by the diffusion model.
This synergistic mechanism effectively accelerates the early-stage convergence process of the SDA-MARL.

In the later training stages, the proposed SDA-MARL exhibits stable convergence curves.
This stability primarily stems from two synergistic mechanisms: first, the accumulation of high-quality samples in the replay buffer under supervised guidance; second, the stabilizing effect of the diffusion model's behavioral cloning loss during policy updates. 
This cloning loss function constrains the DDPG's Actor outputs within a reasonable range, while the diffusion model inherently provides excellent stability and smoothness. Consequently, the auxiliary DDPG network avoids significant fluctuations during updates, ensuring robust convergence performance throughout the later training phases.

\subsubsection{\textbf{Tracking Accuracy}}
In multi-AUV tracking tasks, tracking accuracy serves as a critical metric for evaluating system performance. 
In our evaluations, we establish a tracking accuracy threshold of 0.08 as the performance benchmark. Comprehensive experiments are conducted across four representative scenarios: 2 AUVs tracking 1 target, 4 AUVs tracking 2 targets, 6 AUVs tracking 2 targets, and 8 AUVs tracking 3 targets.
The results in Table~\ref{tab:tracking_accuracy} demonstrate that our SDA-MARL achieves average tracking accuracies of 66.44\%, 68.56\%, 61.38\%, and 61.67\% in the four scenarios, respectively.

\subsubsection{\textbf{Velocity Difference}} 
In multi-AUV collaborative tracking tasks, frequent speed adjustments by AUVs lead to significant increases in propulsion energy consumption, which undermines the long-term operational sustainability of multi-AUV systems. 
To quantitatively evaluate this performance aspect, this paper introduces the velocity difference between AUVs and targets.

Comprehensive experimental results presented in Table~\ref{tab:velocity_mean} demonstrate that our SDA-MARL achieves the lowest mean velocity differences across all four tracking scenarios (2 AUVs tracking 1 target, 4 AUVs tracking 2 targets, 6 AUVs tracking 2 targets, 8 AUVs tracking 3 targets). 
In particular, as shown in Table \ref{tab:velocity_std}, the standard deviation of SDA-MARL is relatively low compared to that of other algorithms. 
These results validate the algorithm's effectiveness in balancing tracking responsiveness and energy conservation while maintaining consistent performance across varying swarm scales and target configurations.
These results indicate that the proposed SDA-MARL aims to enable AUVs to rapidly approach targets during the initial tracking phase, subsequently stabilizing the relative velocity between AUVs and targets at a consistently low level. 
This strategy not only reduces overall system energy consumption but also reflects the stability and convergence characteristics of the control process. 

\subsubsection{\textbf{Path Length}}
Path length serves as a critical indicator of navigation efficiency for AUVs, directly reflecting the AUV's ability to identify optimal trajectories during target pursuit. Well-trained AUVs tend to develop strategies that minimize travel distance while maintaining effective tracking performance. This metric provides quantitative validation of policy optimization quality, where shorter paths indicate superior spatial reasoning and decision-making capabilities in complex underwater environments.

Experimental results in Table~\ref{tab:path_length} demonstrate that the SDA-MARL algorithm consistently achieves the shortest path lengths across all evaluation scenarios compared to the other methods. This performance advantage indicates the proposed SDA-MARL's effectiveness in developing energy-efficient navigation strategies that balance rapid target acquisition with minimal trajectory deviation. The consistent superiority across varying swarm configurations further validates the robustness of our approach in optimizing path planning under different operational complexities.

\begin{figure}[bth]
	\centering
	\includegraphics[width=0.96\linewidth]{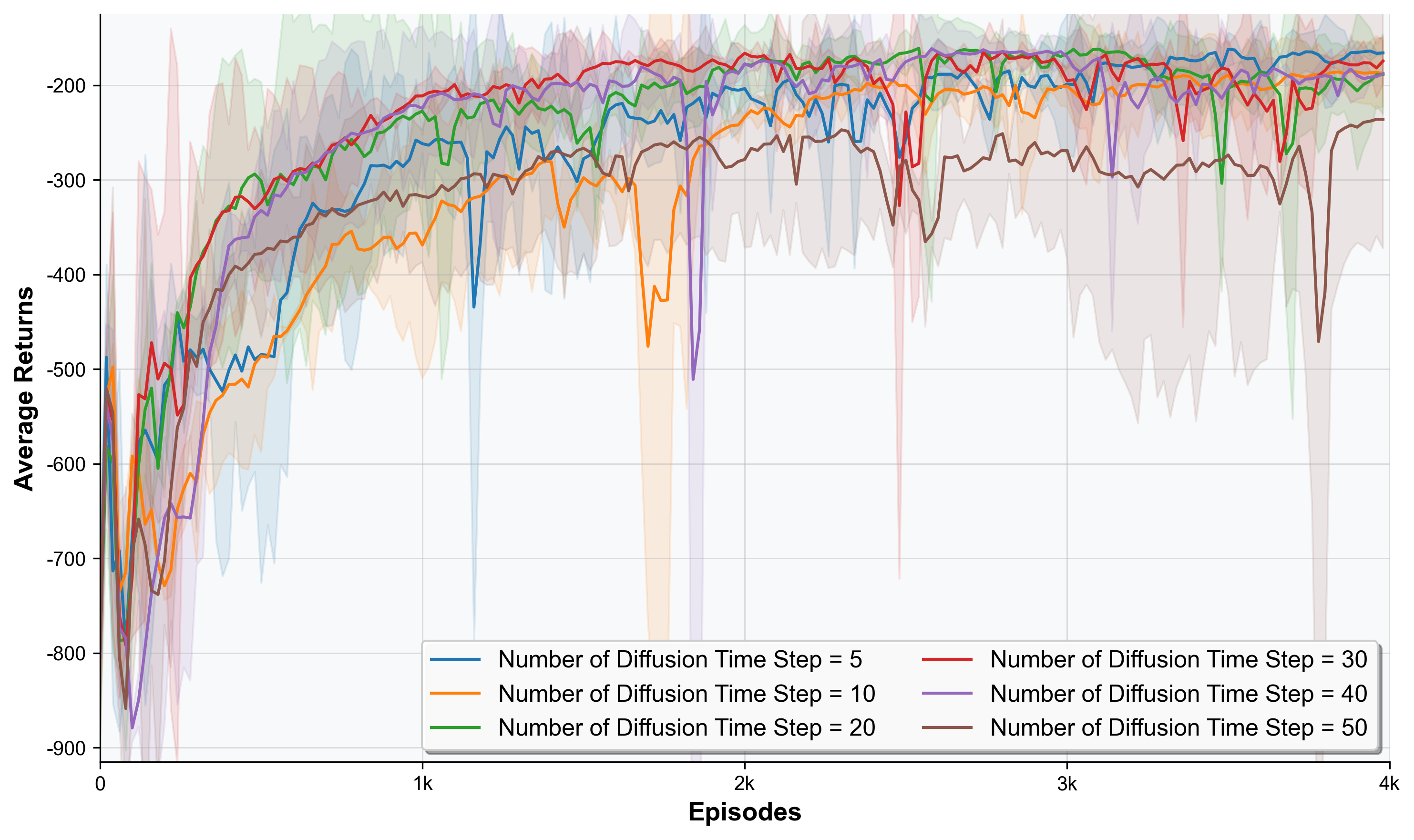}
	\caption{Convergence speed across different numbers of diffusion time steps in SDA-MARL}
	\label{fig4}
\end{figure}

\begin{figure*}[t!]
 	\centering
 	\subfloat[Ablation analysis in 4 AUVs tracking 2 tragets]
 	{\includegraphics[width=0.48\textwidth]{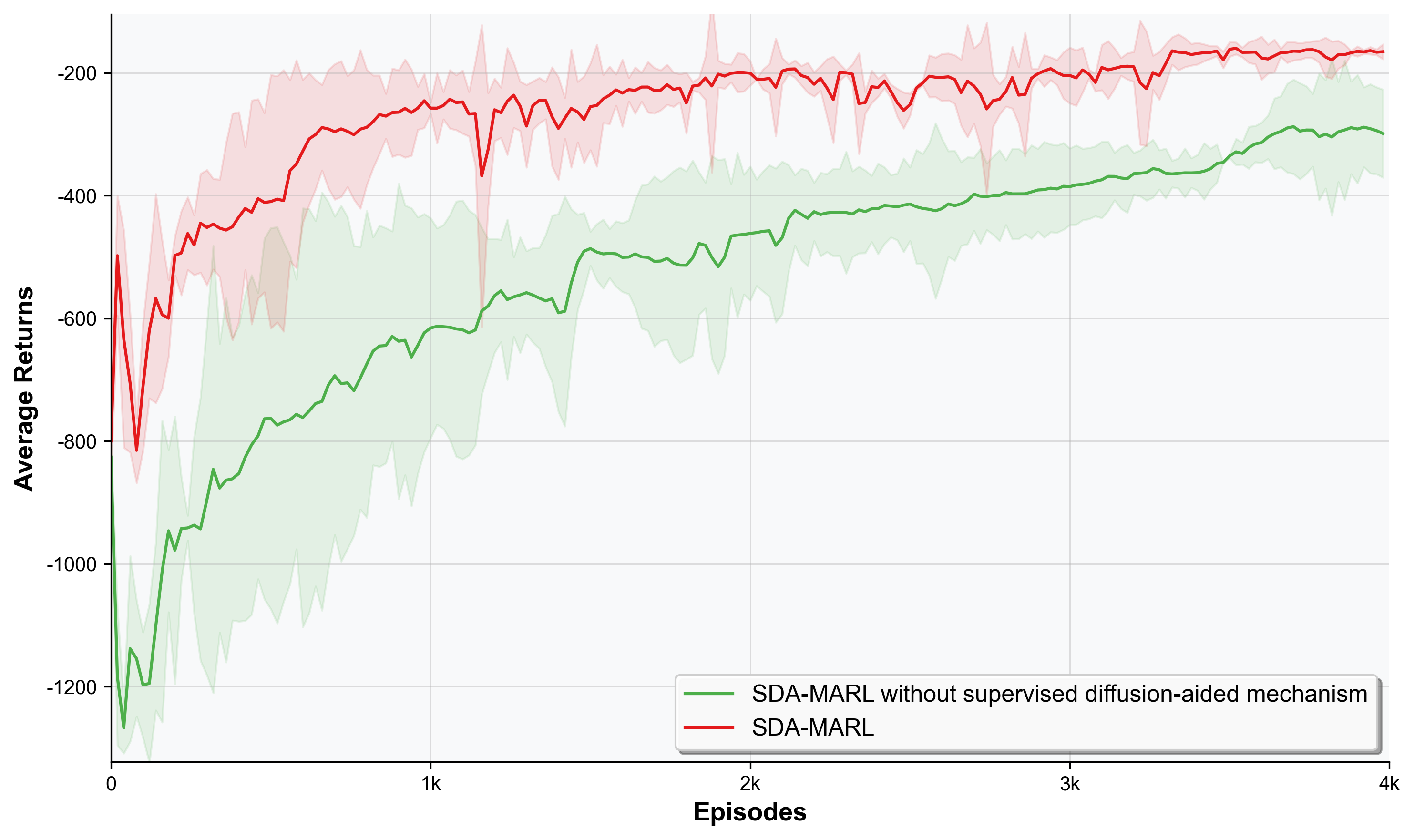}}\hfill
 	\subfloat[Ablation analysis in 8 AUVs tracking 3 tragets]
 	{\includegraphics[width=0.48\textwidth]{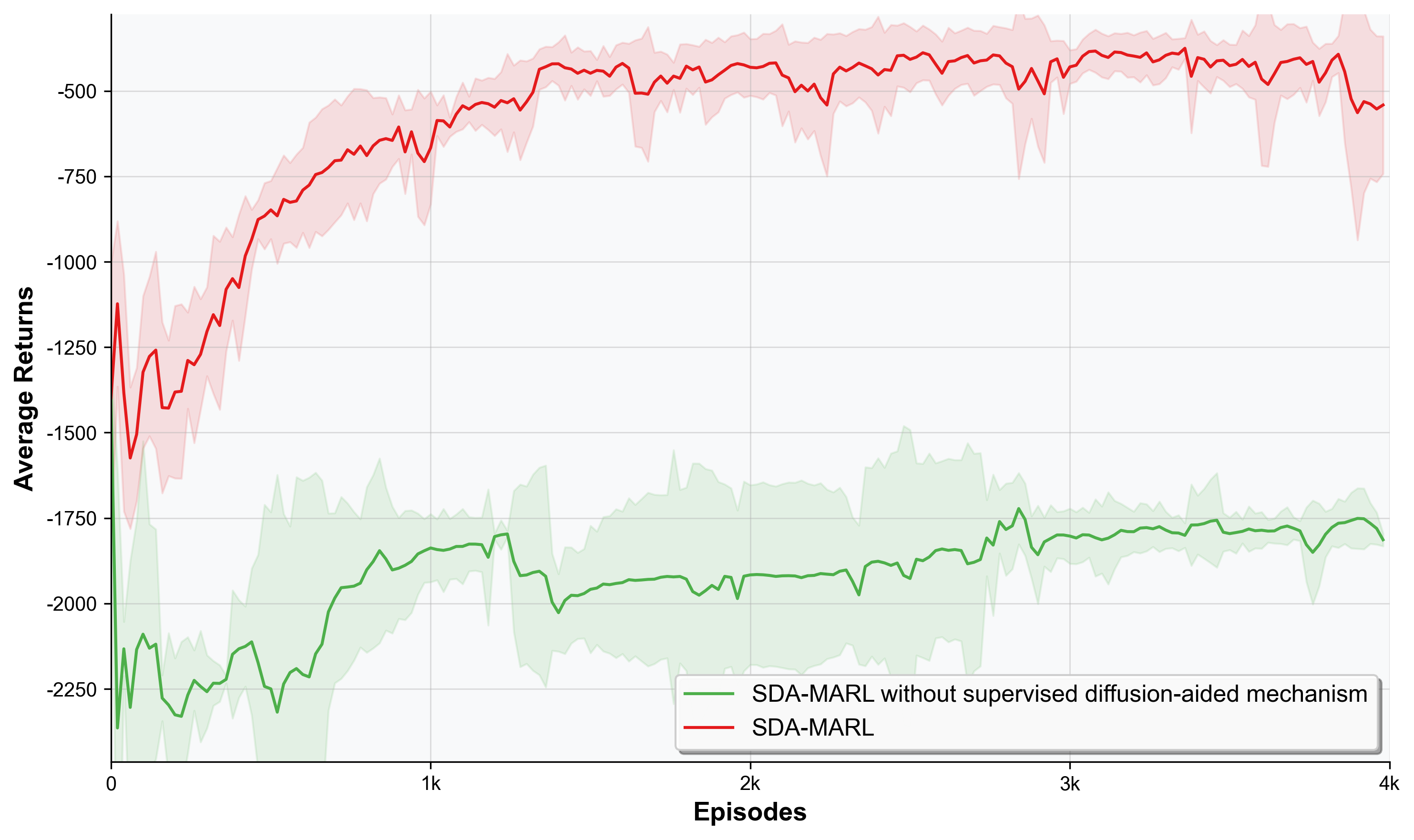}}
 	
 	\caption{Ablation evaluation}
 	\label{fig5}
\end{figure*}

\begin{figure*}[t]
    \centering
    \scriptsize 
    \begin{tabular}{@{}ccc@{}}
        \includegraphics[width=0.31\textwidth]{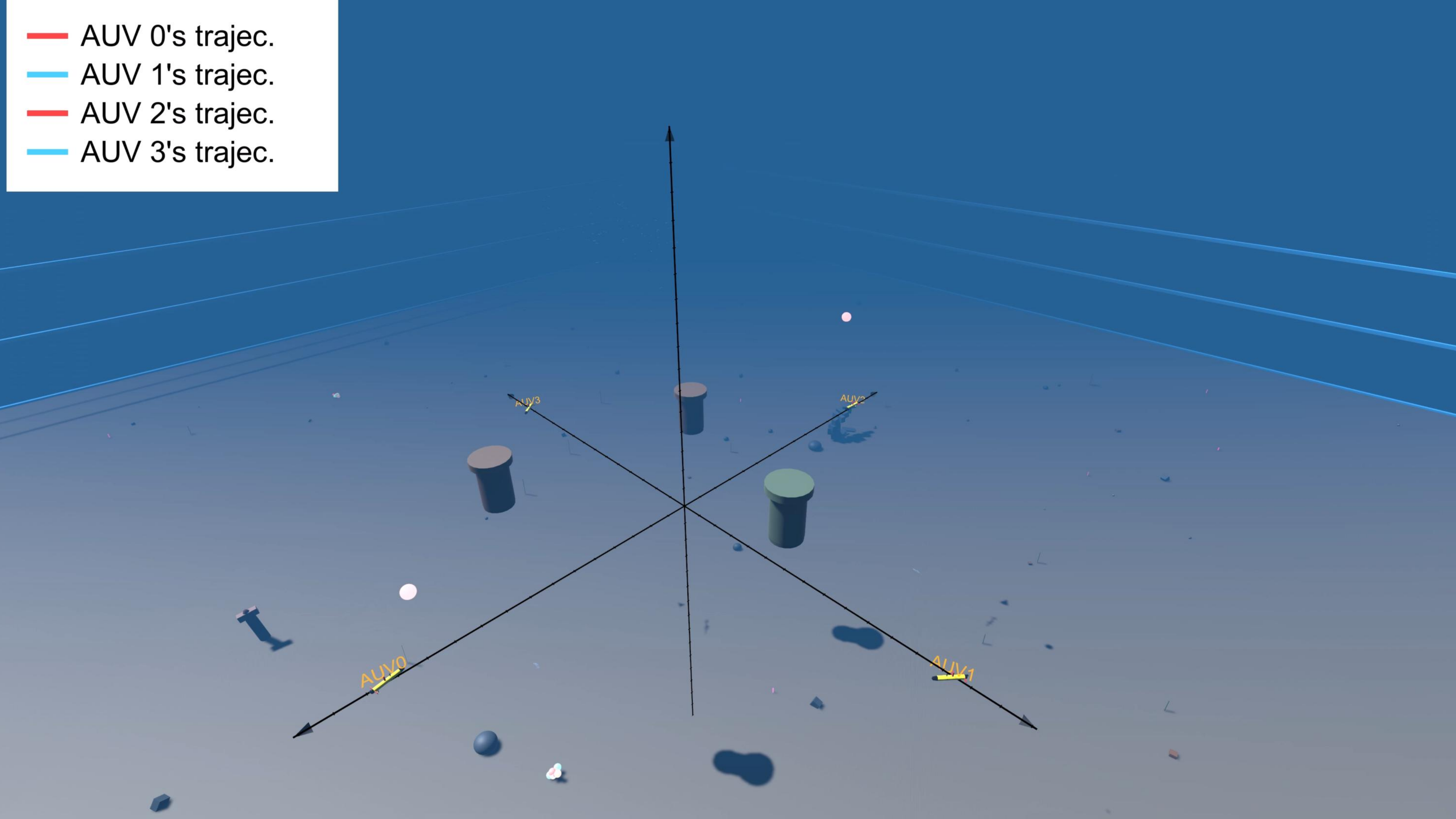} &
        \includegraphics[width=0.31\textwidth]{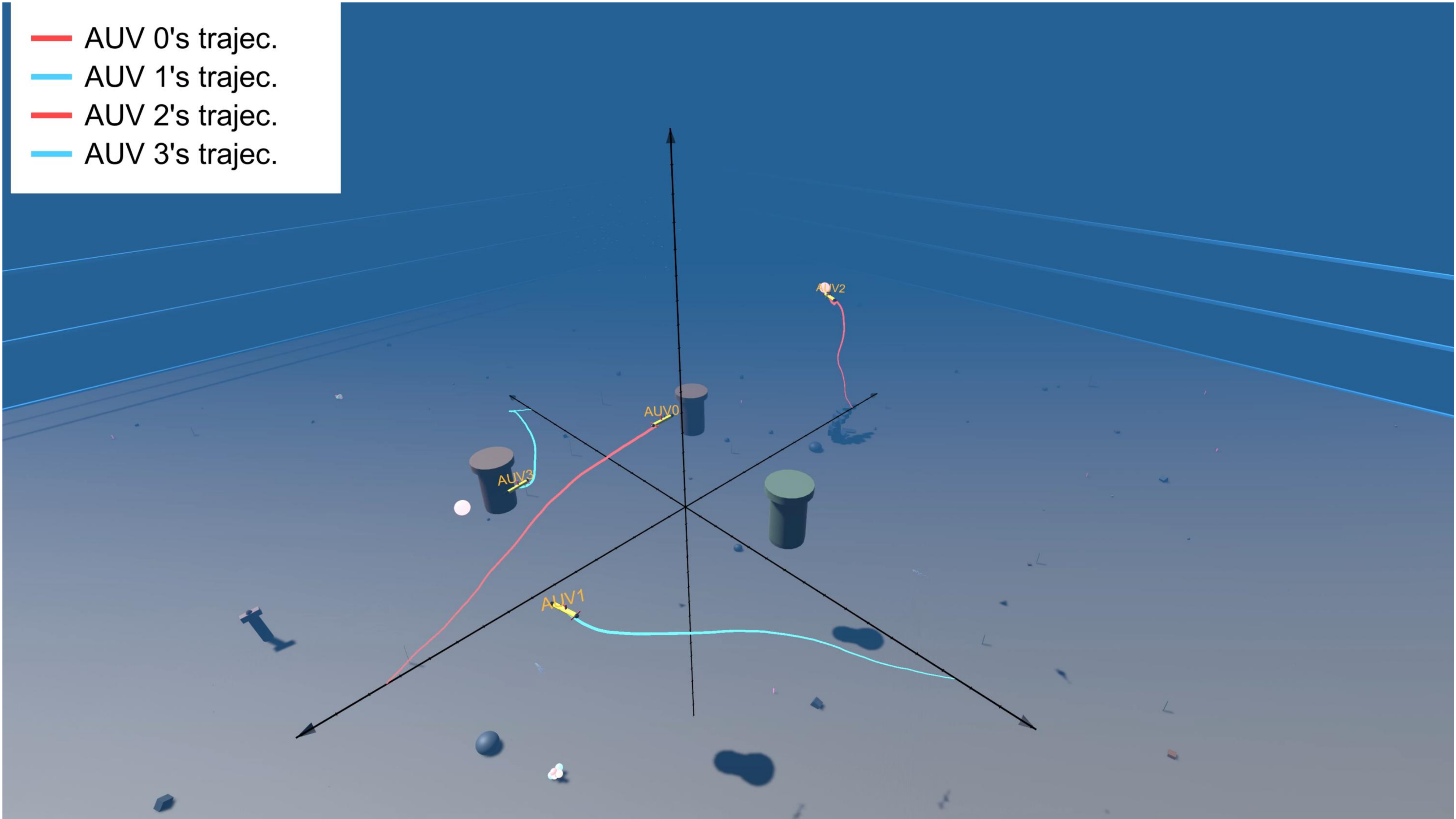} &
        \includegraphics[width=0.31\textwidth]{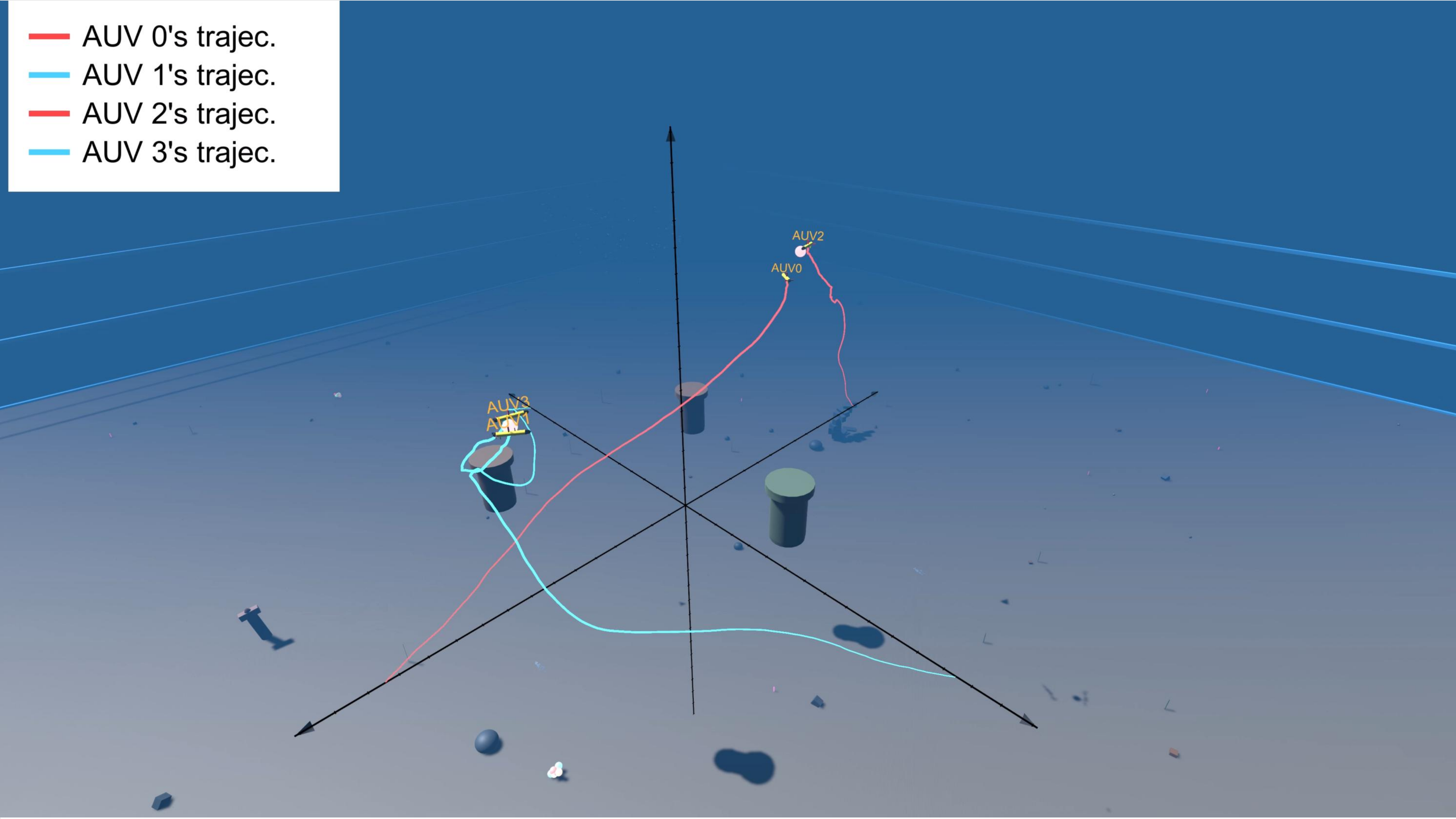} \\
        (a) Initial phase in 4 AUVs tracking 2 targets & (b) Mid-phase in 4 AUVs tracking 2targets & (c) Final phase in 4 AUV tracking 2 targets \\[0.1em]
        
        \includegraphics[width=0.31\textwidth]{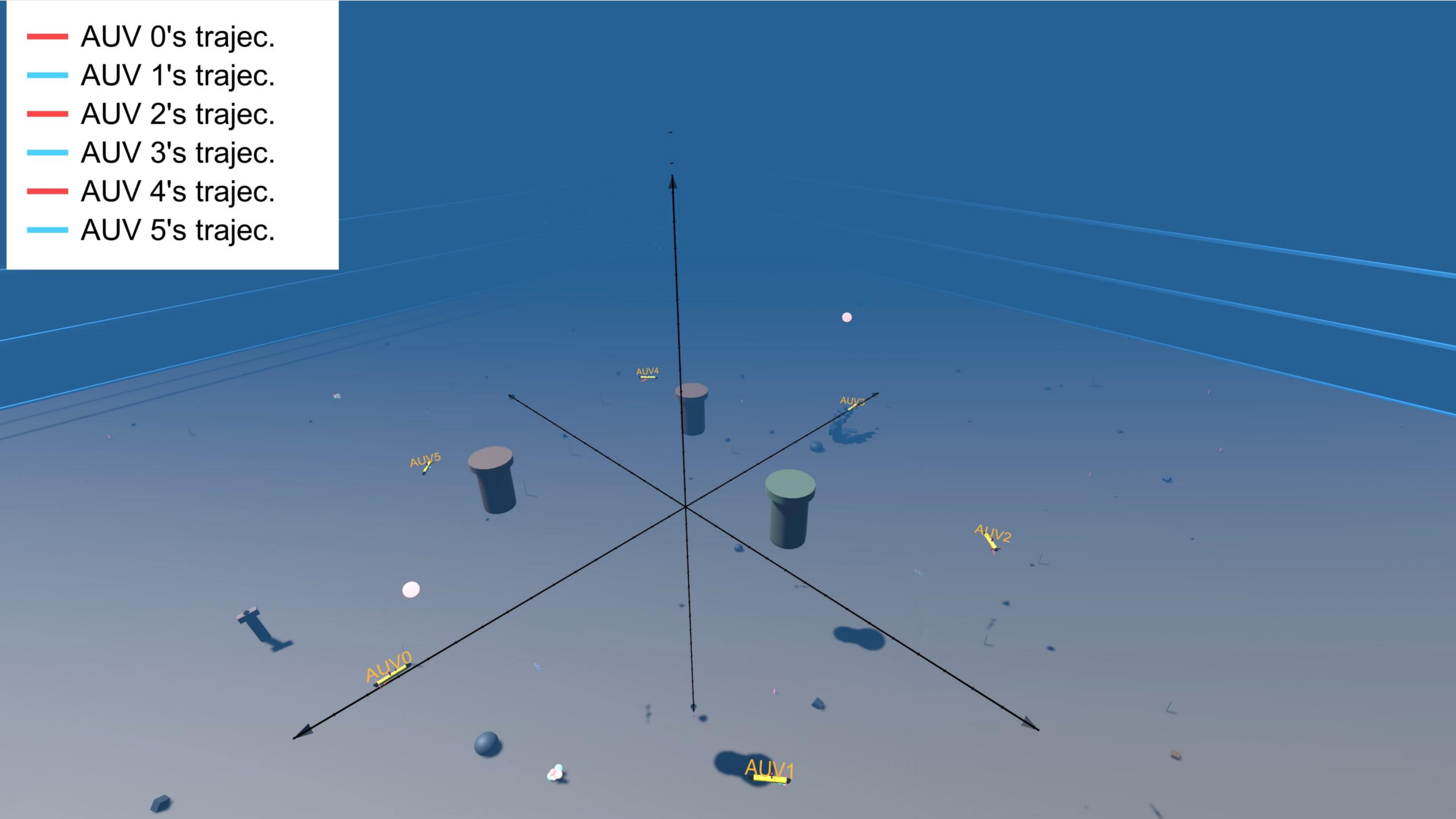} &
        \includegraphics[width=0.31\textwidth]{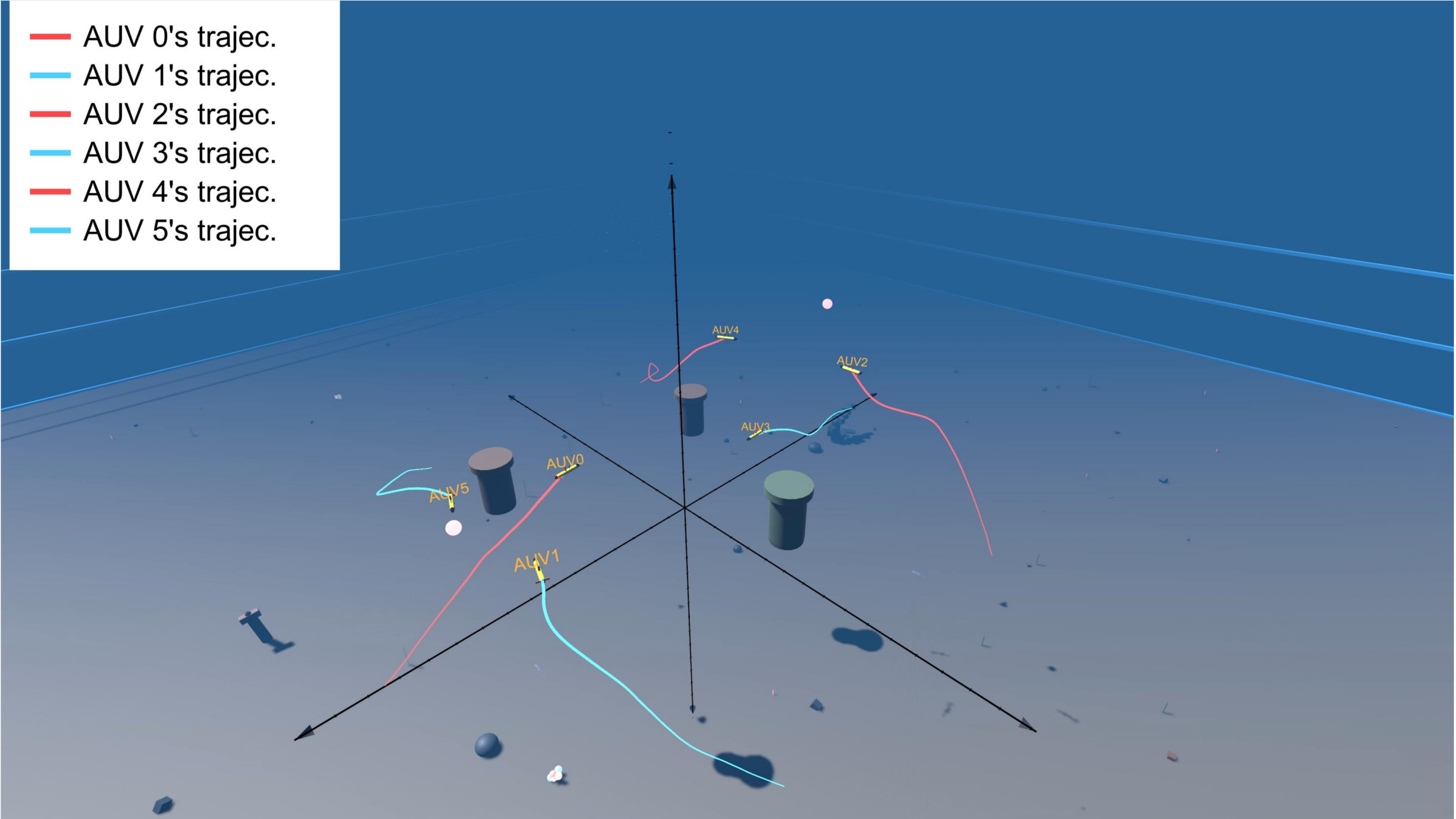} &
        \includegraphics[width=0.31\textwidth]{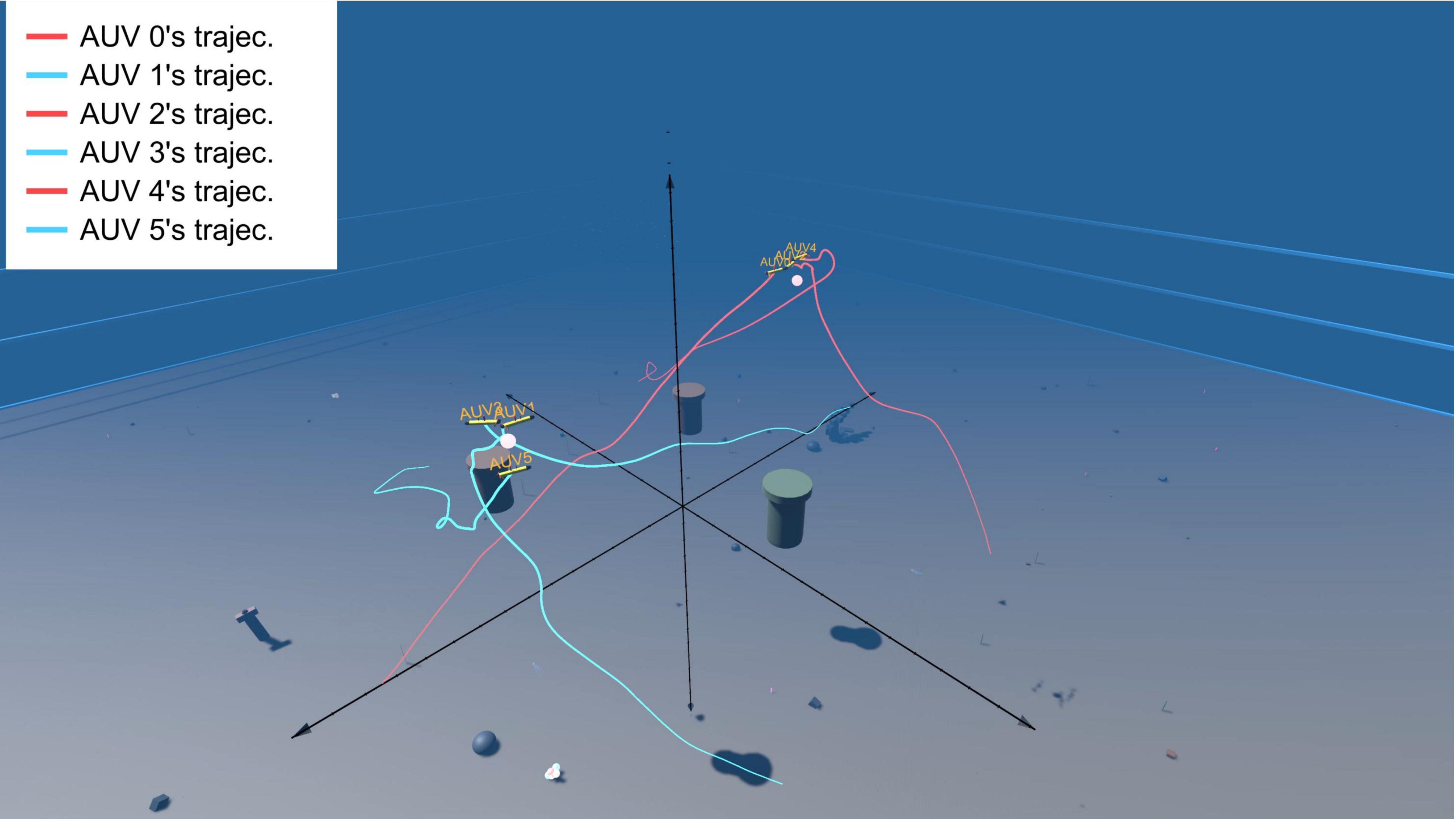} \\
        (d) Initial phase in 6 AUVs track 2 targets & (e) Mid-phase in 6 AUVs tracking 2 targets & (f) Final phase in 6 AUVs tracking 2 targets \\[0.1em]
        
        \includegraphics[width=0.31\textwidth]{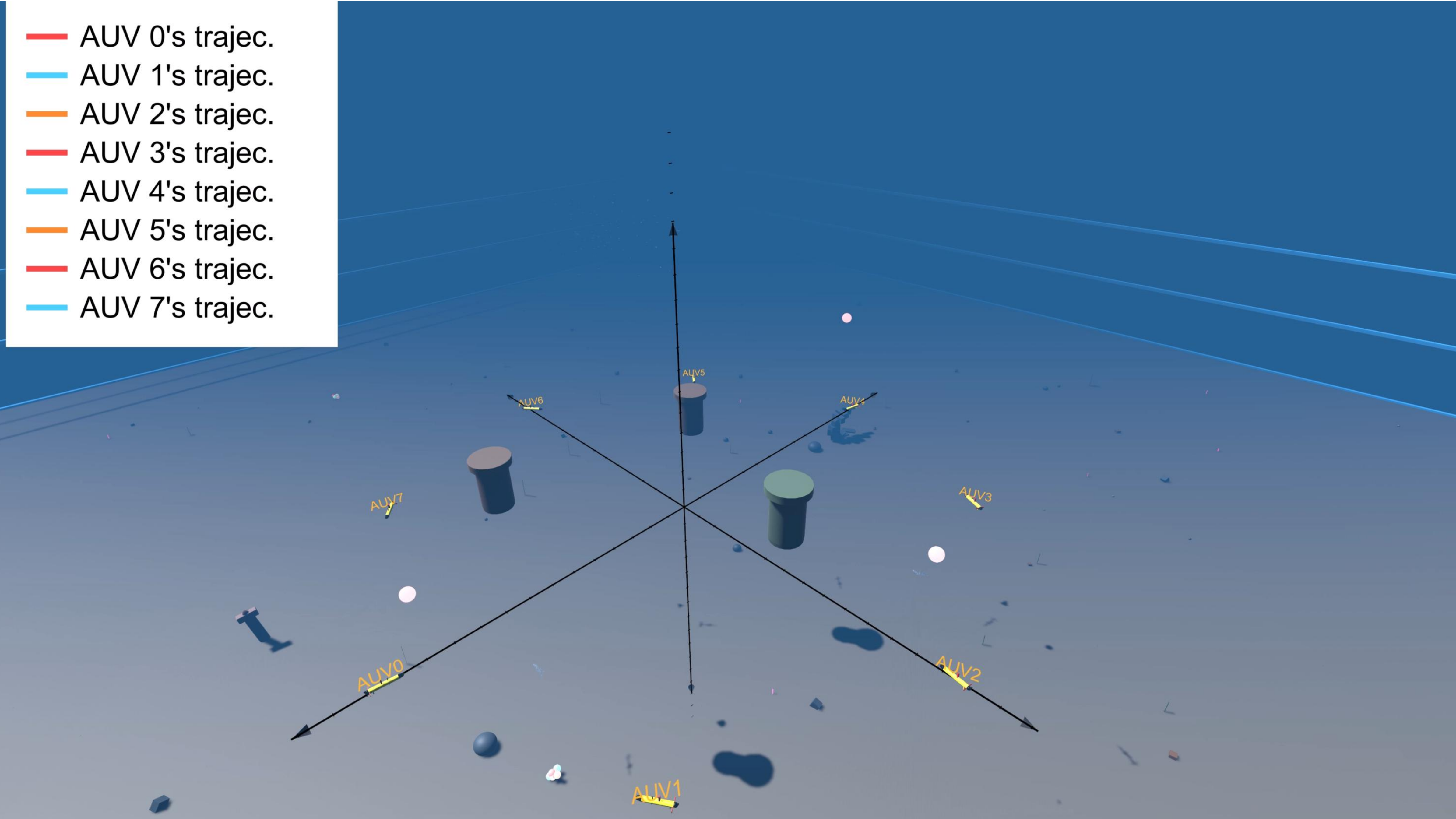} &
        \includegraphics[width=0.31\textwidth]{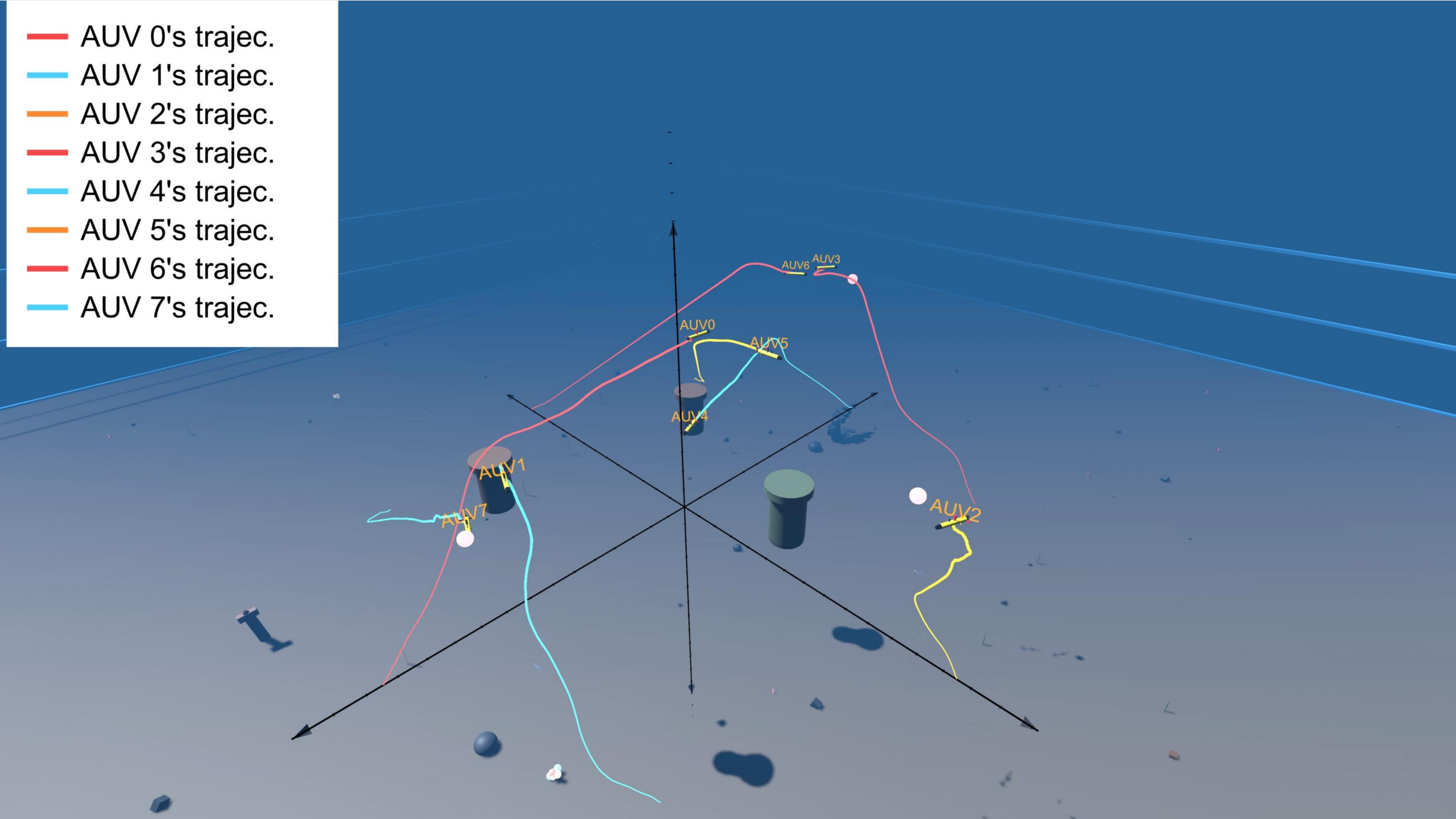} &
        \includegraphics[width=0.31\textwidth]{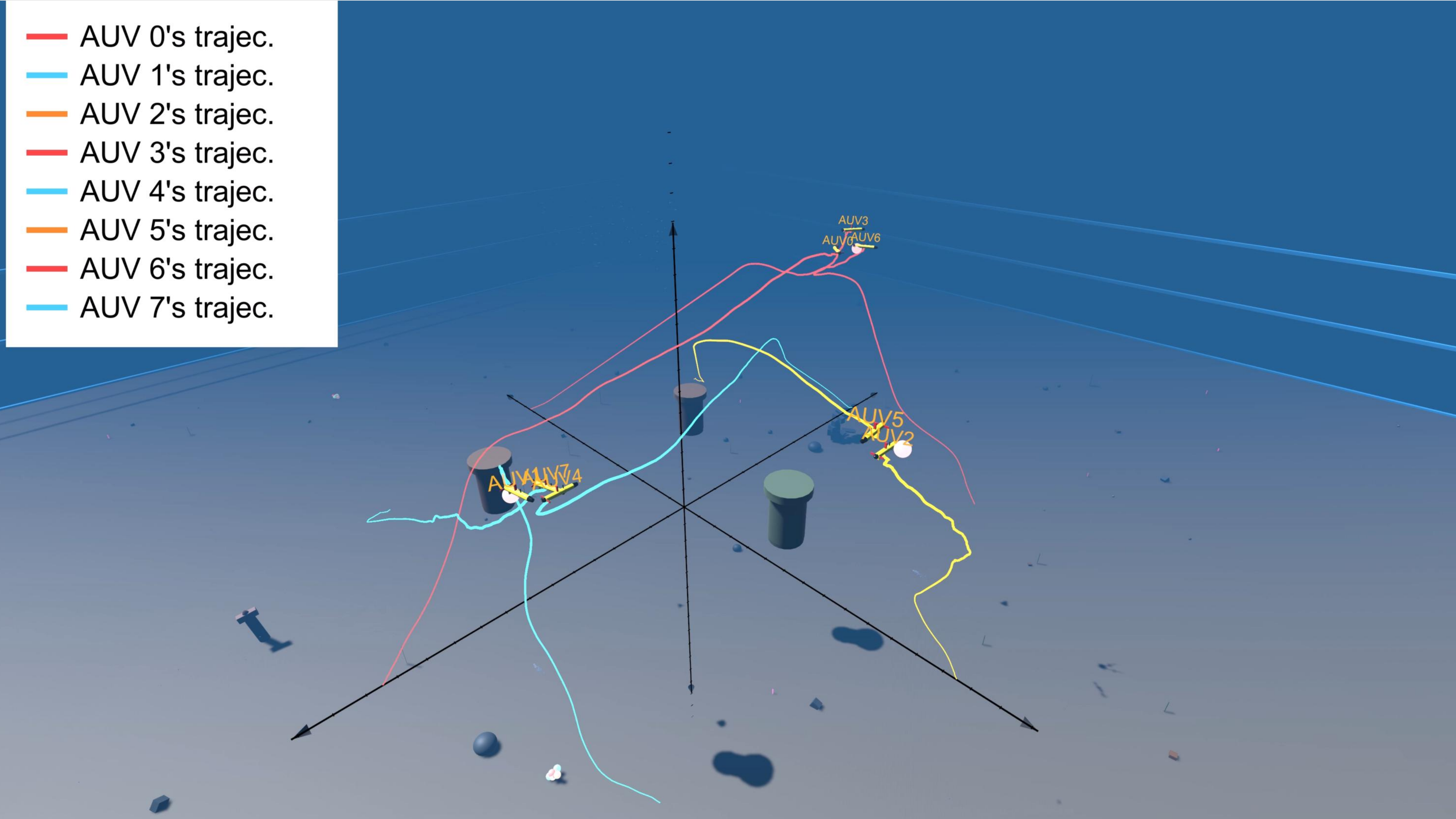} \\
        (g) Initial phase in 8 AUVs tracking 3 targets & (h) Mid-phase in 8 AUVs tracking 3 targets & (i) Final phase in 8 AUVs tracking 3 targets
    \end{tabular}
    \caption{Availability evaluation.}
    \label{fig:multi_track_systematic}
\end{figure*}

\subsubsection{\textbf{Number of Diffusion Time Step}}
Fig. \ref{fig4} presents the average return underwater for different numbers of diffusion time steps ranging from 5 to 50. 
The results in Fig. \ref{fig4} show similar convergence trends under different numbers of diffusion time steps.
Furthermore, as the number of diffusion time steps increases, the convergence speed demonstrates a non-monotonic trend in the medium step range, first rising to a local peak and then declining, which reveals a non-linear dependency between the number of diffusion time steps and system convergence.

\subsubsection{\textbf{Ablation Evaluation}}
Fig. \ref{fig5} compares the proposed SDA-MARL algorithm with an ablated version that removes the diffusion model module, its supervised learning guidance component, and the behavioral cloning loss. Without these components, agents cannot leverage high-quality actions distilled from expert trajectories in the replay buffer, resulting in unstable training due to unmitigated non-stationarity and inefficient exploration in sparse reward underwater environments. This directly increases suboptimal behaviors such as collisions and target loss, reducing accumulated rewards.

The SDA-MARL algorithm achieves higher average rewards and training stability by synergistically integrating diffusion-based action generation, supervised policy guidance, and behavioral cloning. This confirms the essential role of these integrated components in enabling robust coordination for multi-AUV underwater tracking tasks under dynamic conditions.

\subsubsection{\textbf{Availability Evaluation}}

To further evaluate the proposed approach, simulations are conducted in a near-realistic underwater environment developed in Unity3D, with all experiments set in 3D scenes containing dynamic obstacles.

Fig. \ref{fig:multi_track_systematic} illustrates the cooperative tracking process in three typical multi-AUV scenarios. The figure presents the AUVs (yellow mobile entities), the targets (glowing spheres), the oceanic obstacles (screw-shaped objects), and the color-coded trajectories shared by AUVs tracking the same target. The process is chronologically divided into three stages: initial AUV deployment, mid-phase stable tracking, and final consistent trajectory adherence. 

Results show that the proposed algorithm effectively coordinates the multi-AUV system, maintaining tracking within an acceptable error range and thereby demonstrating the availability, adaptability, and robustness of SDA-MARL under varied conditions.



\section{Conclusion}\label{Section:7}

In the paper, we conduct an in-depth investigation into the problem of cooperative tracking of multiple underwater targets by multi-AUV systems. 
We construct a hierarchical AUV MARL architecture integrated with a diffusion model. 
Building upon this architecture, the paper proposes a tracking algorithm based on a proposed SDA-MARL. 
The proposed SDA-MARL innovatively designs a dual-decision architecture combining the diffusion model and DDPG: the policy distribution generated by the diffusion model serves as a supervisory signal, and the DDPG network's policy optimization is guided by a cloning loss, effectively improving the algorithm's convergence speed and tracking stability. 
Furthermore, the proposed SDA-MARL introduces a supervised learning mechanism, dynamically partitioning the experience replay pool based on supervised labels, allowing the dual decision modules (diffusion model and DDPG) to independently optimize network parameters based on dedicated experience subsets, effectively decoupling training interference and accelerating policy convergence.
Experimental results demonstrate that, compared with current mainstream MARL algorithms, the proposed SDA-MARL algorithm exhibits significant advantages in convergence efficiency and other tracking performance metrics.





\appendices


\ifCLASSOPTIONcaptionsoff
  \newpage
\fi



%
\bibliographystyle{IEEEtran}
\bibliography{ref}

\vspace{-5ex}

\begin{IEEEbiography}[{\includegraphics[width=1in,height=1.25in,clip,keepaspectratio]{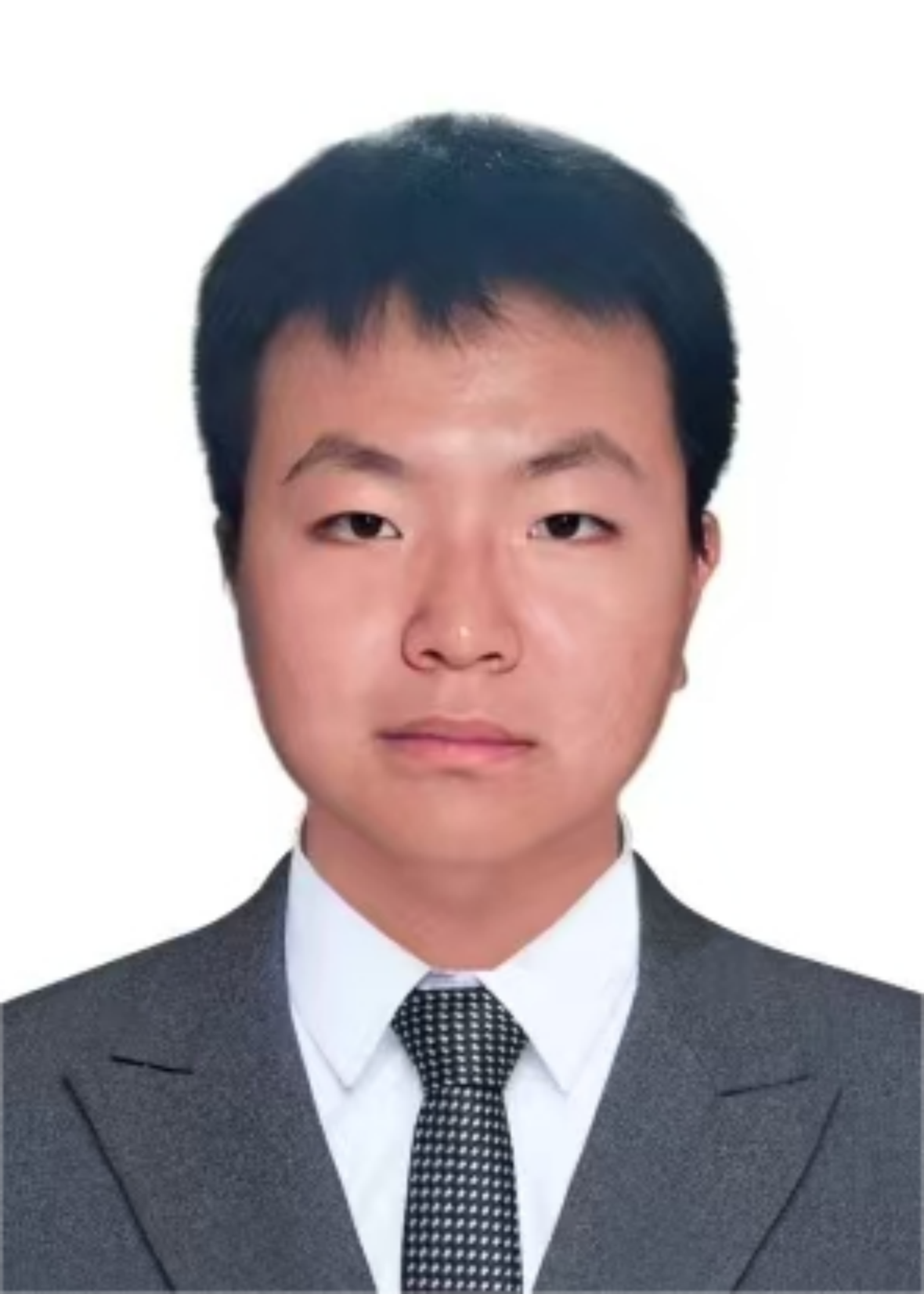}}]{Jiaao Ma}
is currently pursuing a Bachelor's degree at the Software College, Northeastern University, Shenyang, China. His research interests include reinforcement learning, diffusion models, and supervised learning.
\end{IEEEbiography}

\vspace{-5ex}
\begin{IEEEbiography}[{\includegraphics[width=1in,height=1.25in,clip,keepaspectratio]{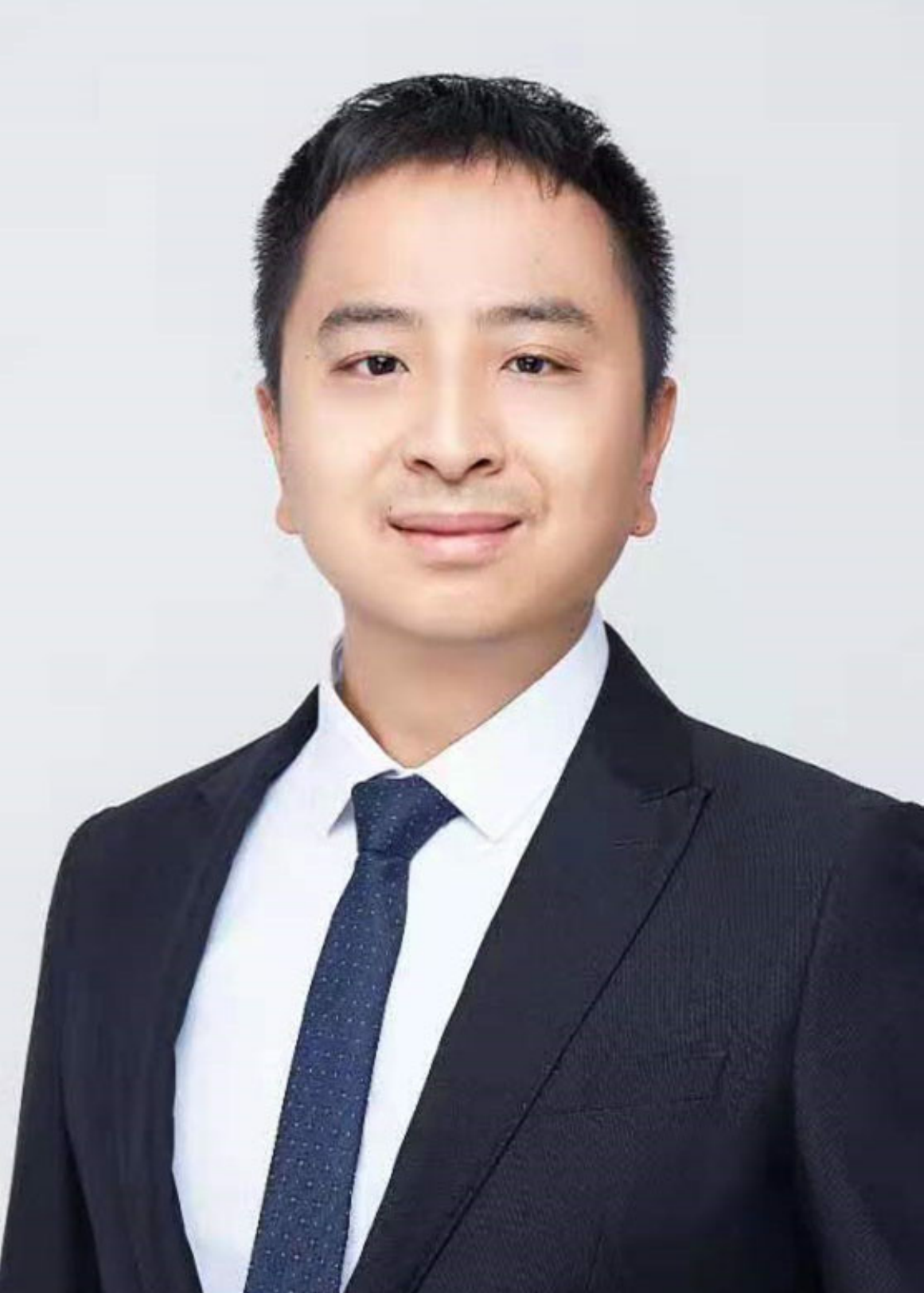}}]{Chuan Lin}
	[S'17, M'20] is currently an associate professor with the Software College, Northeastern University, Shenyang, China.
	He received the B.S. degree in Computer Science and Technology from Liaoning University, Shenyang, China in 2011, the M.S. degree in Computer Science and Technology from Northeastern University, Shenyang, China in 2013, and the Ph.D. degree in computer architecture in 2018.
	From Nove. 2018 to  Nove. 2020, he is a Postdoctoral Researcher with the School of Software, Dalian University of Technology, Dalian, China.
	His research interests include UWSNs, industrial IoT, software-defined networking.
\end{IEEEbiography}
\vspace{-5ex}
\begin{IEEEbiography}[{\includegraphics[width=1in,height=1.25in,clip,keepaspectratio]{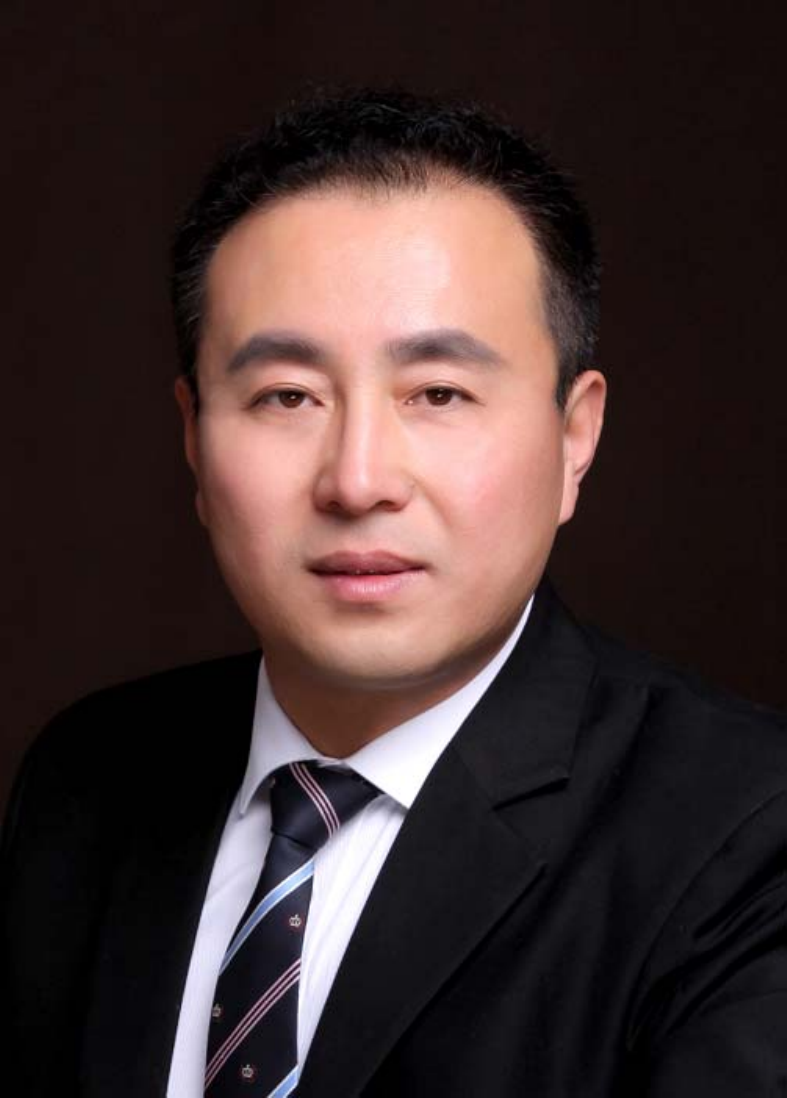}}]{Guangjie Han} [S’03-M’05-SM’18-F’22] is currently a Professor with the Department of Internet of Things Engineering, Hohai University, Changzhou, China. He received his Ph.D. degree from Northeastern University, Shenyang, China, in 2004. In February 2008, he finished his work as a Postdoctoral Researcher with the Department of Computer Science, Chonnam National University, Gwangju, Korea. From October 2010 to October 2011, he was a Visiting Research Scholar with Osaka University, Suita, Japan. From January 2017 to February 2017, he was a Visiting Professor with City University of Hong Kong, China. From July 2017 to July 2020, he was a Distinguished Professor with Dalian University of Technology, China. His current research interests include Internet of Things, Industrial Internet, Machine Learning and Artificial Intelligence, Mobile Computing, Security and Privacy. Dr. Han has over 500 peer-reviewed journal and conference papers, in addition to 160 granted and pending patents. Currently, his H-index is 65 and i10-index is 282 in Google Citation (Google Scholar). The total citation count of his papers raises above 15500+ times. Dr. Han is a Fellow of the UK Institution of Engineering and Technology (FIET). He has served on the Editorial Boards of up to 10 international journals, including the IEEE TII, IEEE TCCN, IEEE TVT, IEEE Systems, etc. He has guest-edited several special issues in IEEE Journals and Magazines, including the IEEE JSAC, IEEE Communications, IEEE Wireless Communications, Computer Networks, etc. Dr. Han has also served as chair of organizing and technical committees in many international conferences. He has been awarded 2020 IEEE Systems Journal Annual Best Paper Award and the 2017-2019 IEEE ACCESS Outstanding Associate Editor Award. He is a Fellow of IEEE.
\end{IEEEbiography}
\vspace{-5ex}
\begin{IEEEbiography}[{\includegraphics[width=1in,height=1.25in,clip,keepaspectratio]{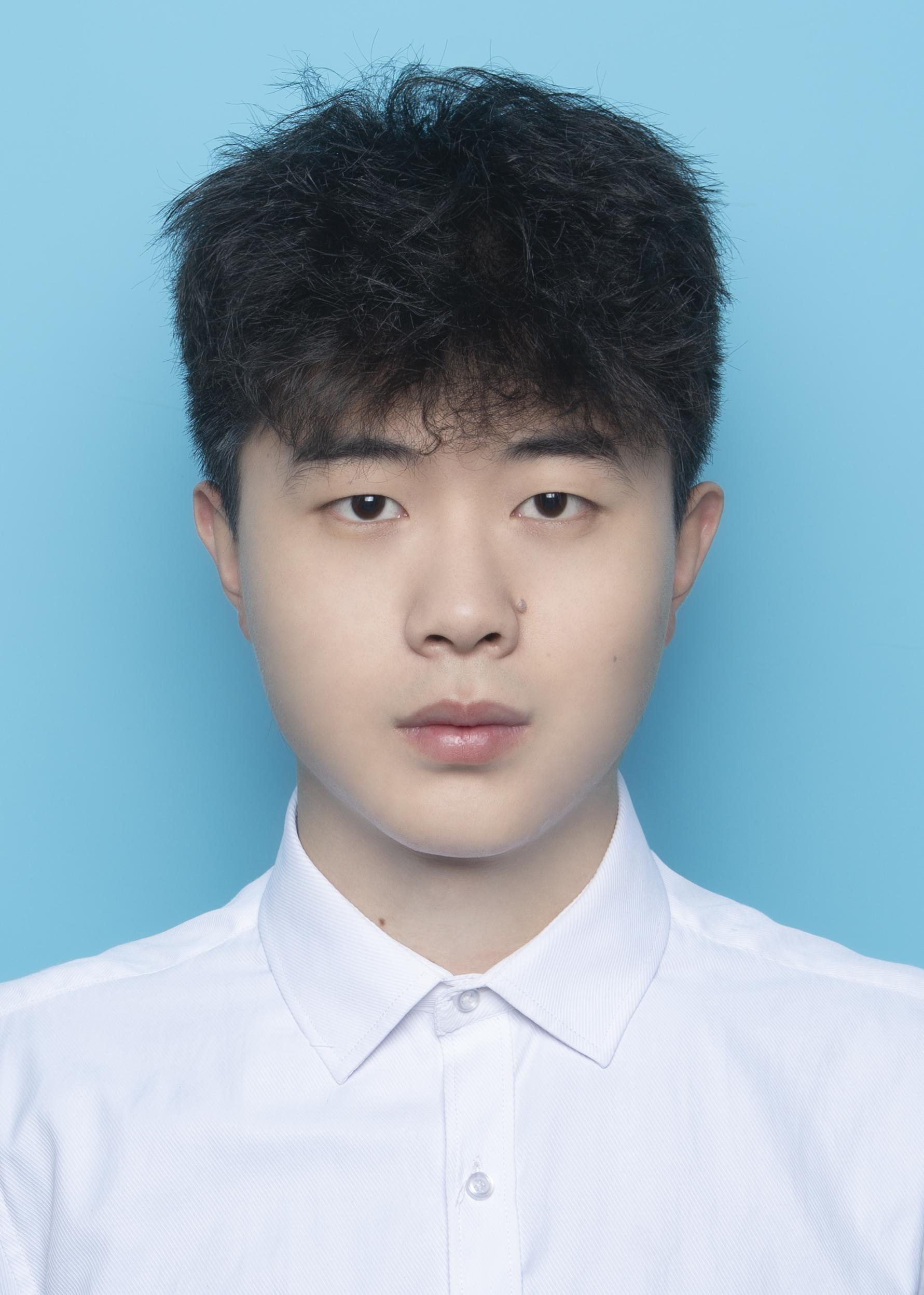}}]{Shengchao Zhu} (Student member, IEEE) received his B.S. degree in Internet of Things Engineering from Hohai University, Changzhou, China, in 2023. He is currently pursuing the Ph.D. degree with the Department of Computer Science and Technology at Hohai University, Nanjing, China. His current research interests include swarm intelligence, swarm ocean, Multi-Agent Reinforcement Learning.
\end{IEEEbiography}
\vspace{-5ex}
\begin{IEEEbiography}[{\includegraphics[width=1in,height=1.25in,clip,keepaspectratio]{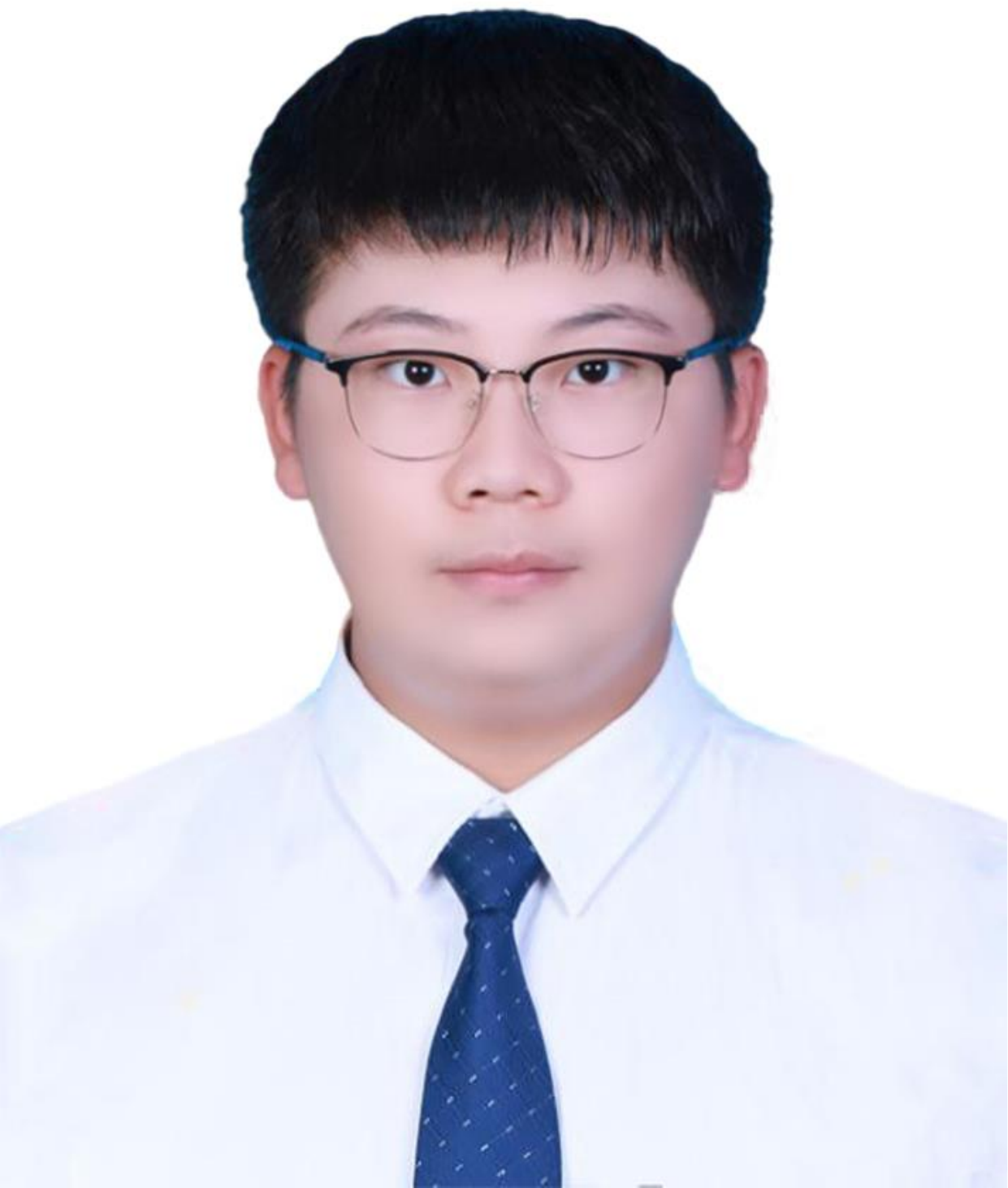}}]{Zhenyu Wang}
is currently pursuing a Bachelor's degree at the Software College, Northeastern University, Shenyang, China. His research interests include reinforcement learning, the Internet of Things, and software-defined networking.
\end{IEEEbiography}
\vspace{-5ex}
\begin{IEEEbiography}[{\includegraphics[width=1in,height=1.25in,clip,keepaspectratio]{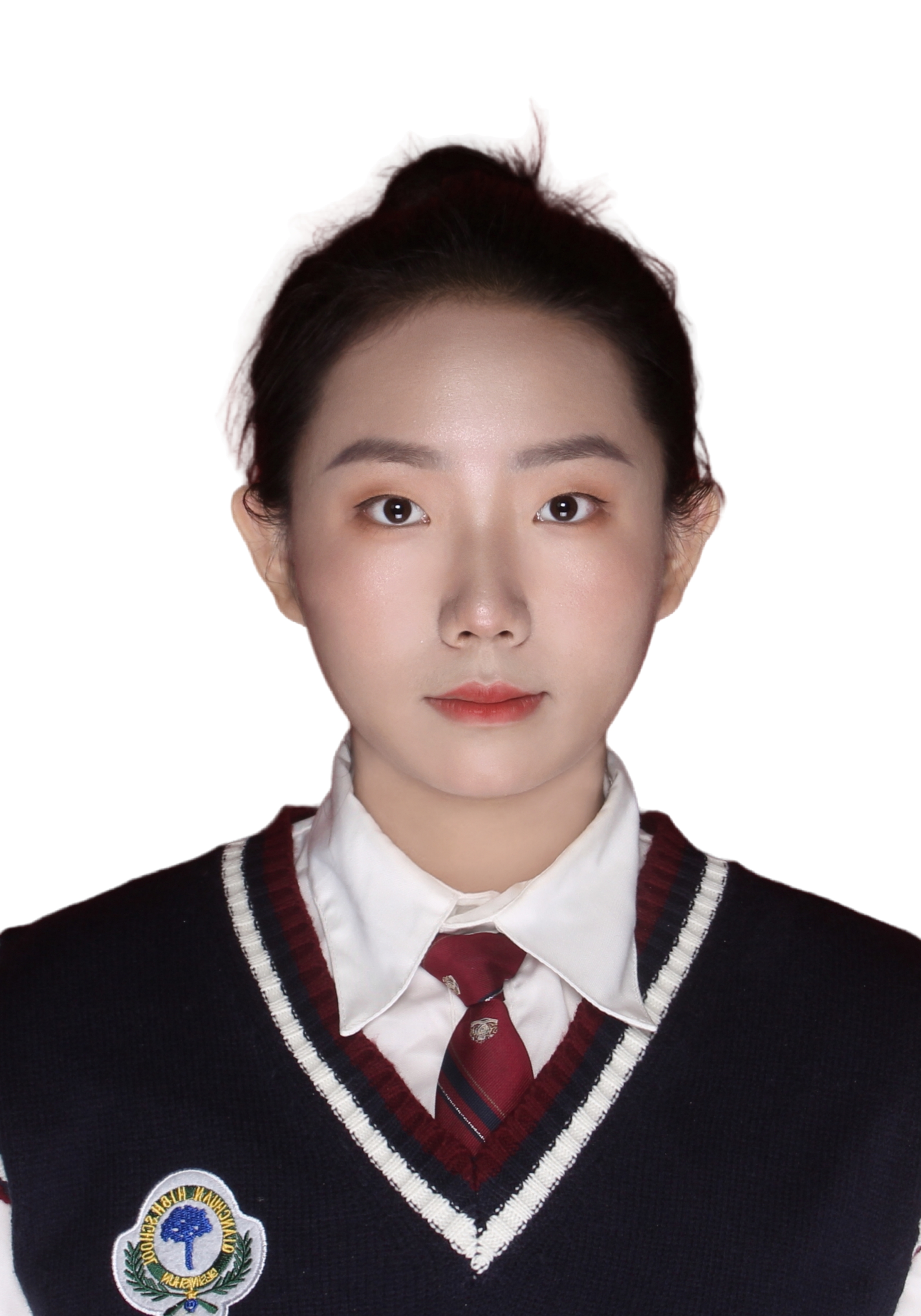}}]{Chen An}
is currently pursuing the B.S. degree at the School of Computer Science and Engineering, Northeastern University, China. Her research interests include affective computing, emotional-expression-related hallucinations in large language models, and model interpretability.
\end{IEEEbiography}
\vspace{-5ex}

\end{document}